\documentclass[
 reprint,
 amsmath,amssymb,
 aps,
]{revtex4-2}

\usepackage{graphicx}
\usepackage{dcolumn}
\usepackage{bm}
\usepackage{slashed}
\usepackage{multirow}
\usepackage{fontsize}
\usepackage{subcaption}
\usepackage[colorlinks=true, linkcolor=red, citecolor=blue, urlcolor=blue]{hyperref}
\begin{document}

\preprint{APS/123-QED}

\title{Scalar and spin polarizabilities of singly heavy baryons within heavy baryon chiral perturbation theory}

\author{Zi-Jun Li$^{1,2,3}$}
  \email{lzijun2024@lzu.edu.cn}
\author{Zhan-Wei Liu$^{1,2,3}$}
  \email{liuzhanwei@lzu.edu.cn}
\author{Ping Chen$^{1,2,3}$}
\affiliation{
 $^1$School of Physical Science and Technology, Lanzhou University, Lanzhou 730000, China\\
 $^2$Research Center for Hadron and CSR Physics, Lanzhou University and Institute of Modern Physics of CAS, Lanzhou 730000, China\\
 $^3$Lanzhou Center for Theoretical Physics, MoE Frontiers Science Center for Rare Isotopes, Key Laboratory of Quantum Theory and Applications of MoE, Key Laboratory of Theoretical Physics of Gansu Province, Gansu Provincial Research Center for Basic Disciplines of Quantum Physics, Lanzhou University, Lanzhou 730000, China
}

\begin{abstract}
We present a systematic study of the scalar electromagnetic and spin polarizabilities of spin-1/2 singly charmed baryons at $\mathcal{O}(p^4)$ within the framework of heavy baryon chiral perturbation theory. Our results show that the higher-order corrections to the electric polarizability are small, while those to the magnetic polarizability are relatively larger due to the small mass splitting of singly charmed baryons and are closely related to transition magnetic moments. Furthermore, we find that the spin polarizabilities of singly charmed baryons, except for $\gamma_{M1M1}$, are much smaller than those of the nucleons. We have also calculated the polarizabilities for singly bottom baryons, with the results showing generally larger values than those of singly charmed baryons.
\end{abstract}

\maketitle

\section{\label{sec:1} INTRODUCTION }

In classical electromagnetism, the electromagnetic polarizabilities $\alpha_E$ and $\beta_M$ are defined as the linear response coefficients relating the induced electromagnetic dipole moments to the external quasistatic electromagnetic fields $\bm{E}$ and $\bm{H}$ \cite{Holstein:2013kia}, and the relative potential energies are $U_E = -\frac{1}{2}4 \pi \alpha_E \bm{E}^2$ and $U_M = -\frac{1}{2}4 \pi \beta_M \bm{H}^2$. For hadrons, these polarizabilities can be probed via Compton scattering where the incident photon provides the external electromagnetic fields. In addition to the scalar electromagnetic polarizabilities $\alpha_E$, $\beta_M$, there are four spin polarizabilities for a hadron with spin 1/2. Two of them $\gamma_{E1E1}$, $\gamma_{M1M1}$ are dipole polarizabilities, and the other two $\gamma_{E1M2}$, $\gamma_{M1E2}$ are quadrupole polarizabilities. These spin polarizabilities provide valuable information about the internal structure of hadrons.

Numerous experimental studies to date have measured the electromagnetic polarizabilities for nucleons \cite{Federspiel:1991yd, Zieger:1992jq, MacGibbon:1995in, deLeon:2001dnx, Blanpied:2001ae,Koester:1995nx, Kolb:2000ix, Lundin:2002jy}. By measuring the nucleon spin structure functions $g_{1,2}$ at JLab, the generalized spin polarizabilities could be indirectly extracted \cite{Chen:2010qc, JeffersonLabHallAg2p:2022qap}. The A2 Collaboration reported the first direct measurement of all four proton spin polarizabilities in 2015 \cite{A2:2014iky}. In 2022, MAMI achieved the separate extraction of all four proton spin polarizabilities for the first time \cite{Mornacchi:2022cln}. The scalar and spin polarizabilities of baryons have also been studied with various theoretical approaches such as Nambu--Jona--Lasinio model \cite{Nikolov:1993ty, Schumacher:2006cy, Schumacher:2007xr}, Skyrme model \cite{Scoccola:1995tf, Gobbi:1995de, Scoccola:1996kh, Tanushi:1997ur}, dispersion relation \cite{Holstein:1994tw, Drechsel:1999rf, Holstein:1999uu, Schumacher:2009xt, Schumacher:2011gs, Schumacher:2013hu}, QCD sum rule \cite{Nishikawa:1997sm}, QCD string theory \cite{Kruglov:1999hc, Kruglov:2001cy}, phenomenological model MAID \cite{Drechsel:2000ct}, perturbative chiral quark model \cite{Dong:2005kt}, covariant Faddeev approach \cite{Eichmann:2016tbi}, and the  established chiral perturbation theory \cite{Bernard:1991rq, Bernard:1991ru, Bernard:1993bg, Bernard:1993ry, Butler:1992ci, Babusci:1996jr, Beane:2004ra, Choudhury:2007qiz, Deshmukh:2017ciw, Lensky:2009uv, Aleksejevs:2013cda, Lensky:2014efa, HillerBlin:2015vgy, Lensky:2015awa, Thurmann:2020mog, Bernard:1995dp, Hemmert:1997tj, Gellas:2000mx, Hemmert:1996rw, Kambor:1997ns, Ji:1999sv, VijayaKumar:2000pv, Griesshammer:2015ahu, Kondratyuk:2001qu, Gasparyan:2011yw, VijayaKumar:2011uw, Bernard:2012hb, HillerBlin:2015vgy, Lensky:2015awa}. Reviews of experimental and theoretical works on nucleon polarizabilities can be found in Refs.~\cite{Holstein:2013kia, Schumacher:2005an, Hagelstein:2015egb, Hagelstein:2020vog,Fonvieille:2019eyf, Sparveris:2024kjz}.

Polarizabilities are well-defined observables characterizing the low-energy electromagnetic response and internal structure of a particle. Spin polarizabilities encode correlations between the spin and the time dependence and spatial gradients of external electromagnetic fields. Spin polarizabilities have also been investigated for octet hyperons \cite{VijayaKumar:2011uw, HillerBlin:2015vgy} and the deuteron \cite{Ji:2003ia}. Particles with different spins have different spin polarizabilities. The vector polarizabilities of $^{87}\mathrm{Sr} $ \cite{westergaard2011lattice} and $^{171}\mathrm{Yb}$ \cite{lemke2009spin} have been measured, and this polarizability mediates dispersive light-spin interactions which were employed to realize a quantum nondemolition measurement of the collective spin of an ensemble of $^{133}\mathrm{Cs}$ atoms with spin-4 \cite{Jia:2023mze}. 

Experimentally, there seems to be no direct measurement of the polarizabilities of singly heavy baryons at present. The greatest difficulty in measuring singly charmed baryons is their extremely short lifetime, on the order of $10^{-13}$ seconds, making them nearly impossible to measure. However, in Ref. \cite{Aiola:2020yam} feasible measurement methods for the electromagnetic properties of singly charmed baryons have been proposed. By bombarding a tungsten target with the 7 TeV proton beam to produce charmed baryons with momenta at the TeV scale whose average flight distance would be extended to the centimeter level due to the relativistic effects, one can then use the bent crystals with extremely strong intrinsic electric fields to induce the spin precession. The measurement would be achieved by accumulating data over a period of two years \cite{Cesare:2024gie, Cesare:2025eqt, Libralon:2026iyc}.

For lattice calculations, both the background field method and the emerging four-point correlation function method have attempted to calculate the spin polarizabilities. Though the background field method is fully capable of calculating spin polarizabilities \cite{Detmold:2006vu}, scalar polarizabilities are mainly focused on due to technical difficulties such as excessive background noise that prevents the extraction of the small signals of spin polarizabilities \cite{Endrodi:2024cqn, Bignell:2020xkf, Kabelitz:2024aye}. On the other hand, the four-point correlation function method was proposed precisely to solve the difficulty in extracting spin polarizability signals \cite{Engelhardt:2011qq}, and the scalar polarizabilities of light mesons and light baryons are currently studied to verify its feasibility \cite{Wilcox:2021rtt, Wang:2023omf}. Furthermore, one can simulate the Mellin moments of the baryon spin structure functions $g_{1,2}$ with lattice QCD \cite{Burger:2021knd}, which are helpful for predicting the generalized spin polarizabilities.

We will study the scalar and spin polarizabilities of singly heavy baryons within heavy baryon chiral perturbation theory (HBChPT) in this work \cite{Weinberg:1978kz, Gasser:1983yg, Gasser:1984gg, Jenkins:1990jv, Bernard:1992qa}. The small mass of pion ensures that the photon interactions remain significant at low energies, and consequently long-range chiral contributions from the pion cloud still critically influence the polarizabilities  \cite{Schumacher:2005an, Hagelstein:2015egb, Hagelstein:2020vog, Thomas:2001kw}. The $\mathcal{O}(p^3)$ and $\mathcal{O}(p^4)$ HBChPT calculations of the nucleon electromagnetic polarizabilities confirmed that the long-range pion cloud provides important contribution \cite{Bernard:1991rq, Bernard:1991ru, Bernard:1993bg, Bernard:1993ry}. Subsequently, many theoretical calculations of the baryon electromagnetic polarizabilities were performed within the frameworks of both HBChPT \cite{Butler:1992ci, Babusci:1996jr, Beane:2004ra, Choudhury:2007qiz, Deshmukh:2017ciw} and covariant baryon chiral perturbation theory \cite{Lensky:2009uv, Aleksejevs:2013cda, Lensky:2014efa, HillerBlin:2015vgy, Lensky:2015awa, Thurmann:2020mog}.

The scalar polarizabilities for singly heavy baryons have been studied up to $\mathcal{O}(p^3)$ \cite{Chen:2024xks, Wen:2025xed}. Drawing on previous experience with the nucleon case \cite{Bernard:1993bg, Bernard:1993ry, Beane:2004ra, Ji:1999sv, VijayaKumar:2000pv, Griesshammer:2015ahu}, the $\mathcal{O}(p^4)$ corrections to both the electromagnetic and spin polarizabilities are non-negligible. Moreover, the spin polarizabilities are also important to characterize the internal spin structures of singly heavy baryons, and convergence issues may arise specifically for the spin polarizabilities. Therefore, we include the $\mathcal{O}(p^4)$ corrections to achieve more accurate predictions and to assess the convergence behavior of HBChPT in the context of singly heavy baryons.

This paper is organized as follows. In Sec.~\ref{sec:2}, we present the framework for the scalar and spin polarizabilities via Compton scattering. In Sec.~\ref{sec:3} we provide the analytical results for the singly charmed and  bottom baryons. The numerical results and discussion are given in Sec.~\ref{sec:4}. A short summary follows in Sec.~\ref{sec:5}.

\section{\label{sec:2} FRAMEWORK}

A singly charmed (bottom) baryon consists of one charm (bottom) quark and two light quarks. In SU(3) flavor symmetry, based on the composition of the light quarks, the heavy baryons can be categorized into the symmetric flavor sextet $6_f$ and the antisymmetric flavor antitriplet $\bar{3}_f$. According to the flavors of the light quarks, sextet $6_f$ and antitriplet $\bar{3}_f$ singly charmed (bottom) baryons can be represented by $3\times 3$ matrices \cite{Yan:1992gz}:
{\fontsize{9.3}{11}\selectfont
\begin{equation}
\begin{aligned}
    \mathcal{B}_{\bar{3},c} = 
    \begin{pmatrix}
        0 & \Lambda^+_c & \Xi^+_c \\
        -\Lambda^+_c & 0 & \Xi^0_c \\
        -\Xi^+_c & -\Xi^0_c & 0 \\
    \end{pmatrix},~
    \mathcal{B}_{6,c}^{(*\mu)} = 
    \begin{pmatrix}
        \Sigma^{++}_c & \frac{\Sigma^+_c}{\sqrt{2}} & \frac{\Xi^{\prime +}_c}{\sqrt{2}} \\
        \frac{\Sigma^+_c}{\sqrt{2}} & \Sigma^0_c & \frac{\Xi^{\prime 0}_c}{\sqrt{2}} \\
        \frac{\Xi^{\prime +}_c}{\sqrt{2}} & \frac{\Xi^{\prime 0}_c}{\sqrt{2}} & \Omega^0_c \\
    \end{pmatrix}^{(*\mu)}\!\!\!\!\!,
\end{aligned}
\label{eq1}
\end{equation}
\begin{equation}
\begin{aligned}
    \mathcal{B}_{\bar{3},b} = 
    \begin{pmatrix}
        0 & \Lambda^0_b & \Xi^0_b \\
        -\Lambda^0_b & 0 & \Xi^-_b \\
        -\Xi^0_b & -\Xi^-_b & 0 \\
    \end{pmatrix},~
    \mathcal{B}^{(*\mu)}_{6,b} = 
    \begin{pmatrix}
        \Sigma^{+}_b & \frac{\Sigma^0_b}{\sqrt{2}} & \frac{\Xi^{\prime 0}_b}{\sqrt{2}} \\
        \frac{\Sigma^0_b}{\sqrt{2}} & \Sigma^-_b & \frac{\Xi^{\prime -}_b}{\sqrt{2}} \\
        \frac{\Xi^{\prime 0}_b}{\sqrt{2}} & \frac{\Xi^{\prime -}_b}{\sqrt{2}} & \Omega^-_b \\
    \end{pmatrix}^{(*\mu)}\!\!\!\!\!.
\end{aligned}
\label{eq2}
\end{equation}}


\subsection{Compton scattering tensor}

The effective Hamiltonian of baryons in external electromagnetic fields can be written in the following form:
\begin{equation}
\begin{aligned}
    H =& H_0 - \frac{1}{2} 4 \pi \alpha_E \bm{E}^2 -\frac{1}{2}4 \pi \beta_M \bm{H}^2 \\
    & - \frac{1}{2}4\pi \left(\gamma_{E1E1}\bm{\sigma}\cdot \bm{E} \times \dot{\bm{E}} + \gamma_{M1M1}\bm{\sigma}\cdot \bm{H} \times \dot{\bm{H}}\right. \\
    &\left. + 2\gamma_{E1M2}\sigma^i E^j H_{ij} - 2\gamma_{M1E2}\sigma^i H^j E_{ij} \right) + \cdots ;
\end{aligned}
\label{eq4}
\end{equation}
where $E_{ij} = \frac{1}{2}\left(\nabla_i E_j + \nabla_j E_i\right)$ and $H_{ij} = \frac{1}{2}\left(\nabla_i H_j + \nabla_j H_i\right)$ represent the electromagnetic field gradients. 

The low-energy theorem prescribes the expansion of the Compton amplitude $\mathcal{M}$ in powers of the photon energy $\omega$  \cite{Low:1954kd, Gell-Mann:1954wra}. The terms which include the polarizabilities are given below (higher-order terms are omitted):
\begin{equation}
\begin{aligned}
    \mathcal{M} =& A_1(\omega,\theta) \bm{\epsilon}'\cdot \bm{\epsilon} + A_2(\omega,\theta) \bm{\epsilon}'\cdot \hat{\bm{k}} \bm{\epsilon} \cdot \hat{\bm{k}}' \\
    & + A_3(\omega,\theta) i \bm{\sigma}\cdot \left(\bm{\epsilon}'\times \bm{\epsilon} \right) \\
    & + A_4(\omega,\theta) i \bm{\sigma}\cdot \left(\hat{\bm{k}}'\times \hat{\bm{k}} \right) \bm{\epsilon}'\cdot \bm{\epsilon} \\
    & + A_5(\omega,\theta) i \bm{\sigma}\cdot \left[\left(\bm{\epsilon}'\times \hat{\bm{k}} \right) \bm{\epsilon}\cdot \hat{\bm{k}}' - \left(\bm{\epsilon}\times \hat{\bm{k}}' \right) \bm{\epsilon}'\cdot \hat{\bm{k}}\right] \\
    & + A_6(\omega,\theta) i \bm{\sigma}\cdot \left[\left(\bm{\epsilon}'\times \hat{\bm{k}}' \right) \bm{\epsilon}\cdot \hat{\bm{k}}' - \left(\bm{\epsilon}\times \hat{\bm{k}} \right) \bm{\epsilon}'\cdot \hat{\bm{k}}\right], 
\end{aligned}
\label{eq5}
\end{equation}
where the energy of photon is $\omega = v\cdot k$, $v=(1,0,0,0)$ and $\theta$ is the photon scattering angle. The unit vectors along the photon momenta are denoted by $\hat{\bm{k}}$ and $\hat{\bm{k}}'$ for the incident and outgoing photons, respectively.

We compute the Compton scattering amplitude $\mathcal{M}$ and then extract the polarizabilities by differentiating $A_i$. The scalar polarizabilities are determined by the functions $A_1$ and $A_2$ in Eq.~(\ref{eq5}), while the spin polarizabilities are determined by $A_3$--$A_6$ \cite{Hemmert:1997tj}:
\begin{equation}
\begin{aligned}
    &\alpha_E + \beta_M = \left.\frac{1}{8\pi}\frac{\partial^2}{\partial \omega^2}A_1(\omega,0)\right|_{\omega = 0}, \\
    &\beta_M = \left.-\frac{1}{8\pi}\frac{\partial^2}{\partial \omega^2}A_2(\omega,0)\right|_{\omega = 0}, \\
    &\gamma_i = \left.\frac{1}{24\pi}\frac{\partial^3}{\partial \omega^3}A_j(\omega,\theta_i)\right|_{\omega = 0}.
\end{aligned}
\label{eq6}
\end{equation}
Here, the index mapping is $i=1,2,3,4$ to $j=3,4,6,5$. We set $\theta_{2,3,4}=0$ and $\theta_{1}=\pi/2$.

The formulation above involves two representations of the spin polarizabilities: the set $\gamma_{1\text{-}4}$, used here for computational convenience, and the set $\gamma_{E1E1}$, $\gamma_{M1M1}$, $\gamma_{M1E2}$, $\gamma_{E1M2}$ adopted by the PDG, which are in fact linear combinations of $\gamma_{1\text{-}4}$ \cite{Holstein:2013kia}:
\begin{equation}
\begin{aligned}
    \gamma_{E1E1}&=-\gamma_1-\gamma_3,\qquad
    &\gamma_{M1M1}&=\gamma_4, \\
    \gamma_{E1M2}&=\gamma_3, \qquad
    &\gamma_{M1E2}&=\gamma_2+\gamma_4.
\end{aligned}
\label{eq7}
\end{equation}

\subsection{Effective Lagrangians in HBChPT}

We now present the Lagrangians in HBChPT framework. The corresponding Lagrangians can be found in Refs.~\cite{Chen:2024xks, Scherer:2002tk, Bernard:1995dp, Ecker:1994gg, Pich:1995bw} and we follow the conventions of Ref.~\cite{Chen:2024xks}. The LO heavy baryon Lagrangian is 
\begin{equation}
\begin{aligned}
    \mathcal{L}^{(1)}_{\mathcal{B} \phi} =& \left\langle\bar{\mathcal{B}}_6 iv\cdot D \mathcal{B}_6\right\rangle - \left\langle\bar{\mathcal{B}}^*_6\left(iv\cdot D - \delta\right)\mathcal{B}^*_6\right\rangle \\
    & + 2g_1\left\langle\bar{\mathcal{B}}_6 S\cdot u \mathcal{B}_6\right\rangle + g_3\left\langle\bar{\mathcal{B}}^*_{6\mu}u^\mu \mathcal{B}_6 + \text{H.c.}\right\rangle \\
    & + 2g_5\left\langle\bar{\mathcal{B}}^{*\mu}_6 S\cdot u \mathcal{B}^*_{6\mu}\right\rangle.
\end{aligned}
\label{eq16}
\end{equation}

Within the HBChPT framework, there is a significant mass difference between the $\mathcal{B}_{\bar{3}}$ and $\mathcal{B}_6^{(*)}$ states. The studies of magnetic moments indicate that they must be decoupled to ensure chiral convergence due to this large mass splitting \cite{Wang:2018gpl, Meng:2018gan, Wang:2018cre}. Moreover, the $\phi \mathcal{B}_{\bar{3}} \mathcal{B}_{\bar{3}}$ vertex is forbidden under exact SU(3) flavor symmetry. If symmetry breaking effects are considered, such contributions would be incorporated in higher-order terms. Therefore, the $\mathcal{B}_{\bar{3}}$ states are not included in our calculation. We focus solely on constructing the Lagrangian for the $\mathcal{B}_6$ and $\mathcal{B}_6^*$ states.

The next-to-leading-order (NLO) heavy baryon Lagrangian is written as

\begin{widetext}
\begin{equation}
\begin{aligned}
    \mathcal{L}^{(2)}_{\mathcal{B} \phi} =& \left\langle\bar{\mathcal{B}}_6\frac{(v\cdot D)^2 - D^2}{2M_6}\mathcal{B}_6\right\rangle - \left\langle\bar{\mathcal{B}}^{*\mu}_6\frac{(v\cdot D)^2 - D^2}{2M_{6^*}}\mathcal{B}^*_{6\mu}\right\rangle - \frac{id_5}{4M_N}\left\langle\bar{\mathcal{B}}_6\left[S^\mu,S^\nu\right]\hat{F}^+_{\mu \nu}\mathcal{B}_6\right\rangle - \frac{id_6}{4M_N}\left\langle\bar{\mathcal{B}}_6\left[S^\mu,S^\nu\right]\mathcal{B}_6\right\rangle \left\langle F^+_{\mu \nu}\right\rangle \\
    & - \frac{ig_1}{M_6}\left\langle\bar{\mathcal{B}}_6 \left\{S\cdot D, v\cdot u\right\}\mathcal{B}_6\right\rangle + \frac{if_6}{4M_N}\left\langle\bar{\mathcal{B}}_6\hat{F}^+_{\mu \nu}S^\nu \mathcal{B}^{*\mu}_6\right\rangle + \text{H.c.} + \frac{if_7}{4M_N}\left\langle\bar{\mathcal{B}}_6S^\nu \mathcal{B}^{*\mu}_6\right\rangle \left\langle F^+_{\mu \nu}\right\rangle+ \text{H.c.} \\
    & - \frac{ig_3}{M_6}\left\langle\bar{\mathcal{B}}^{*\mu}_6 \left\{D_\mu, v\cdot u\right\}\mathcal{B}_6\right\rangle + \text{H.c.} + \frac{if_9}{4M_N}\left\langle\bar{\mathcal{B}}^{*\mu}_6\hat{F}^+_{\mu \nu}\mathcal{B}^{*\nu}_6\right\rangle + \frac{if_{10}}{4M_N}\left\langle\bar{\mathcal{B}}^{*\mu}_6\mathcal{B}^{*\nu}_6\right\rangle \left\langle F^+_{\mu \nu}\right\rangle.
\end{aligned}
\label{eq17}
\end{equation}
\begin{equation}
\begin{aligned}
    \mathcal{L}^{(2)}_{\mathcal{B} \phi \phi} =& c_0 \langle\bar{\mathcal{B}}_6 \mathcal{B}_6\rangle\langle\chi_+\rangle + c_1 \langle\bar{\mathcal{B}}_6 \tilde\chi_+ \mathcal{B}_6\rangle + \left(c_2 - \frac{2g^2_2 + g^2_1}{4M_6}\right)\langle\bar{\mathcal{B}}_6 v\cdot u \enspace v\cdot u \mathcal{B}_6\rangle + \left(c_3 + \frac{2g^2_2 - g^2_1}{4M_6}\right)\bar{\mathcal{B}}_6^{ab} v\cdot u_a^c \enspace v\cdot u_b^d \mathcal{B}_{6,cd} \\
    & + c_4\langle\bar{\mathcal{B}}_6 \mathcal{B}_6\rangle \langle v\cdot u \enspace v\cdot u \rangle + c_5\langle\bar{\mathcal{B}}_6 u\cdot u \mathcal{B}_6\rangle + c_6\bar{\mathcal{B}}^{ab}_6 u_a^c \cdot u_b^d \mathcal{B}_{6,cd} + c_7\langle\bar{\mathcal{B}}_6 \mathcal{B}_6\rangle\langle u\cdot u\rangle.
\end{aligned}
\label{eq17.5}
\end{equation}
\end{widetext}
In the above Lagrangian, $M_6 = 2535 \enspace\text{MeV}$ and $M_{6^*} = 2602 \enspace\text{MeV}$ represent the average baryon masses of the spin-1/2 and spin-3/2 sextets, respectively. $\delta$ denotes the mass splittings between different multiplets $\delta = M_{6^*} - M_{6}$. We follow Ref.~\cite{Chen:2024xks} and use the nucleon mass $M_N$ in the $d_i$ and $f_i$ terms to avoid extra conversion in the numerical program because the corresponding magnetic moments in this manuscript are expressed in units of the nuclear magneton $\mu_N$. We will only use the $c_{0\text{-}4}$ terms in $\mathcal{L}^{(2)}_{\mathcal{B}\phi\phi}$ because the contributions from the $c_{5,6,7}$ terms are linearly related to those from the $c_{2,3,4}$ ones and thus can be absorbed. The $\mathcal{L}^{(4)}_{\mathcal{B} \phi}$ is written as
\begin{equation}
\begin{aligned}
    \mathcal{L}^{(4)}_{\mathcal{B} \phi} =& a_1 \langle\bar{\mathcal{B}}_6 \hat{F}^+_{\mu \nu} \hat{F}^{+\mu \nu} \mathcal{B}_6\rangle + a_2 \langle\bar{\mathcal{B}}_6 \mathcal{B}_6\rangle\langle F^+_{\mu \nu} F^{+\mu \nu} \rangle \\
    & + a_3 \langle\bar{\mathcal{B}}^{ab}_6 \hat{F}^{+c}_{\mu \nu ,a} \hat{F}^{+\mu \nu ,d}_b \mathcal{B}_{6,cd}\rangle \\
    & + a_4 \langle\bar{\mathcal{B}}_6 v^\mu v_\lambda \hat{F}^+_{\mu \nu} \hat{F}^{+\lambda \nu} \mathcal{B}_6\rangle \\
    & + a_5 \langle\bar{\mathcal{B}}_6 \mathcal{B}_6\rangle\langle v^\mu v_\lambda F^+_{\mu \nu} F^{+\lambda \nu} \rangle \\
    & + a_6 \langle\bar{\mathcal{B}}^{ab}_6 v^\mu v_\lambda \hat{F}^{+c}_{\mu \nu ,a} \hat{F}^{+\lambda \nu ,d}_b \mathcal{B}_{6,cd}\rangle.
\end{aligned}
\label{eq18}
\end{equation}

\begin{figure*}[htbp]
    \centering
    \begin{subfigure}[b]{0.19\textwidth}
        \centering
        \includegraphics[page=39, width=1.0\textwidth]{feynman.pdf}
        \caption{($a_2$)}
    \end{subfigure}
    \hfill
    \begin{subfigure}[b]{0.19\textwidth}
        \centering
        \includegraphics[page=1, width=1.0\textwidth]{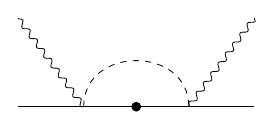}
        \caption{($c_2$)}
    \end{subfigure}
    \hfill
    \begin{subfigure}[b]{0.19\textwidth}
        \centering
        \includegraphics[page=2, width=1.0\textwidth]{feynman.pdf}
        \caption{($c_3$)}
    \end{subfigure}
    \hfill
    \begin{subfigure}[b]{0.19\textwidth}
        \centering
        \includegraphics[page=3, width=1.0\textwidth]{feynman.pdf}
        \caption{($c_4$)}
    \end{subfigure}
    \hfill
    \begin{subfigure}[b]{0.19\textwidth}
        \centering
        \includegraphics[page=5, width=1.0\textwidth]{feynman.pdf}
        \caption{($d_2+e_2$)}
    \end{subfigure}
    \hfill
    \begin{subfigure}[b]{0.19\textwidth}
        \centering
        \includegraphics[page=6, width=1.0\textwidth]{feynman.pdf}
        \caption{($d_3+e_3$)}
    \end{subfigure}
    \hfill
    \begin{subfigure}[b]{0.19\textwidth}
        \centering
        \includegraphics[page=7, width=1.0\textwidth]{feynman.pdf}
        \caption{($d_4+e_4$)}
    \end{subfigure}
    \hfill
    \begin{subfigure}[b]{0.19\textwidth}
        \centering
        \includegraphics[page=8, width=1.0\textwidth]{feynman.pdf}
        \caption{($d_5+e_5$)}
    \end{subfigure}
    \hfill
    \begin{subfigure}[b]{0.19\textwidth}
        \centering
        \includegraphics[page=9, width=1.0\textwidth]{feynman.pdf}
        \caption{($f_2$)}
    \end{subfigure}
    \hfill
    \begin{subfigure}[b]{0.19\textwidth}
        \centering
        \includegraphics[page=10, width=1.0\textwidth]{feynman.pdf}
        \caption{($f_3$)}
    \end{subfigure}
    \hfill
    \begin{subfigure}[b]{0.19\textwidth}
        \centering
        \includegraphics[page=11, width=1.0\textwidth]{feynman.pdf}
        \caption{($f_4$)}
    \end{subfigure}
    \hfill
    \begin{subfigure}[b]{0.19\textwidth}
        \centering
        \includegraphics[page=12, width=1.0\textwidth]{feynman.pdf}
        \caption{($g_2$)}
    \end{subfigure}
    \hfill
    \begin{subfigure}[b]{0.19\textwidth}
        \centering
        \includegraphics[page=13, width=1.0\textwidth]{feynman.pdf}
        \caption{($g_3$)}
    \end{subfigure}
    \hfill
    \begin{subfigure}[b]{0.19\textwidth}
        \centering
        \includegraphics[page=35, width=1.0\textwidth]{feynman.pdf}
        \caption{($h_2$)}
    \end{subfigure}
    \hfill
    \begin{subfigure}[b]{0.19\textwidth}
        \centering
        \includegraphics[page=36, width=1.0\textwidth]{feynman.pdf}
        \caption{($h_3$)}
    \end{subfigure}
    \hfill
    \begin{subfigure}[b]{0.19\textwidth}
        \centering
        \includegraphics[page=37, width=1.0\textwidth]{feynman.pdf}
        \caption{($h_4$)}
    \end{subfigure}
    \hfill
    \begin{subfigure}[b]{0.19\textwidth}
        \centering
        \includegraphics[page=38, width=1.0\textwidth]{feynman.pdf}
        \caption{($h_5$)}
    \end{subfigure}
    \hfill
    \begin{subfigure}[b]{0.19\textwidth}
        \centering
        \includegraphics[page=14, width=1.0\textwidth]{feynman.pdf}
        \caption{($c'_2$)}
    \end{subfigure}
    \hfill
    \begin{subfigure}[b]{0.19\textwidth}
        \centering
        \includegraphics[page=15, width=1.0\textwidth]{feynman.pdf}
        \caption{($c'_3$)}
    \end{subfigure}
    \hfill
    \begin{subfigure}[b]{0.19\textwidth}
        \centering
        \includegraphics[page=16, width=1.0\textwidth]{feynman.pdf}
        \caption{($c'_4$)}
    \end{subfigure}
    \hfill
    \begin{subfigure}[b]{0.19\textwidth}
        \centering
        \includegraphics[page=17, width=1.0\textwidth]{feynman.pdf}
        \caption{($c''_4$)}
    \end{subfigure}
    \hfill
    \begin{subfigure}[b]{0.19\textwidth}
        \centering
        \includegraphics[page=18, width=1.0\textwidth]{feynman.pdf}
        \caption{($c'''_4$)}
    \end{subfigure}
    \hfill
    \begin{subfigure}[b]{0.19\textwidth}
        \centering
        \includegraphics[page=22, width=1.0\textwidth]{feynman.pdf}
        \caption{($d'_2+e'_2$)}
    \end{subfigure}
    \hfill
    \begin{subfigure}[b]{0.19\textwidth}
        \centering
        \includegraphics[page=23, width=1.0\textwidth]{feynman.pdf}
        \caption{($d'_3+e'_3$)}
    \end{subfigure}
    \hfill
    \begin{subfigure}[b]{0.19\textwidth}
        \centering
        \includegraphics[page=24, width=1.0\textwidth]{feynman.pdf}
        \caption{($d'_4+e'_4$)}
    \end{subfigure}
    \hfill
    \begin{subfigure}[b]{0.19\textwidth}
        \centering
        \includegraphics[page=25, width=1.0\textwidth]{feynman.pdf}
        \caption{($d'_5+e'_5$)}
    \end{subfigure}
    \hfill
    \begin{subfigure}[b]{0.19\textwidth}
        \centering
        \includegraphics[page=26, width=1.0\textwidth]{feynman.pdf}
        \caption{($d''_5+e''_5$)}
    \end{subfigure}
    \hfill
    \begin{subfigure}[b]{0.19\textwidth}
        \centering
        \includegraphics[page=27, width=1.0\textwidth]{feynman.pdf}
        \caption{($d'''_5+e'''_5$)}
    \end{subfigure}
    \hfill
    \begin{subfigure}[b]{0.19\textwidth}
        \centering
        \includegraphics[page=28, width=1.0\textwidth]{feynman.pdf}
        \caption{($f'_2$)}
    \end{subfigure}
    \hfill
    \begin{subfigure}[b]{0.19\textwidth}
        \centering
        \includegraphics[page=29, width=1.0\textwidth]{feynman.pdf}
        \caption{($f'_3$)}
    \end{subfigure}
    \hfill
    \begin{subfigure}[b]{0.19\textwidth}
        \centering
        \includegraphics[page=30, width=1.0\textwidth]{feynman.pdf}
        \caption{($f'_4$)}
    \end{subfigure}
    \hfill
    \hspace*{\fill}
    \begin{subfigure}[b]{0.19\textwidth}
        \centering
        \includegraphics[page=31, width=1.0\textwidth]{feynman.pdf}
        \caption{($f''_4$)}
    \end{subfigure}
    \hfill
    \begin{subfigure}[b]{0.19\textwidth}
        \centering
        \includegraphics[page=32, width=1.0\textwidth]{feynman.pdf}
        \caption{($f'''_4$)}
    \end{subfigure}
    \hfill
    \begin{subfigure}[b]{0.19\textwidth}
        \centering
        \includegraphics[page=33, width=1.0\textwidth]{feynman.pdf}
        \caption{($g'_2$)}
    \end{subfigure}
    \hfill
    \begin{subfigure}[b]{0.19\textwidth}
        \centering
        \includegraphics[page=34, width=1.0\textwidth]{feynman.pdf}
        \caption{($g'_3$)}
    \end{subfigure}
    \hspace*{\fill}
    \captionsetup{justification=raggedright, singlelinecheck=false}
    \caption{The contact term and loop diagrams at $\mathcal{O}(p^4)$. The solid dot denotes a vertex or propagator from $\mathcal{L}^{(2)}$. The empty dot denotes a vertex from $\mathcal{L}^{(4)}$. The single solid lines and double solid lines represent spin-1/2 and spin-3/2 baryons, respectively. The crossed and symmetric diagrams are not shown.}
\label{fig1}
\end{figure*}

We employ the Lagrangians in Eqs.~(\ref{eq16})--(\ref{eq18}) to derive the Feynman rules. This work considers Feynman diagrams at $\mathcal{O}(p^3)$ \cite{Chen:2024xks} as well as those at $\mathcal{O}(p^4)$. The $\mathcal{O}(p^4)$ diagrams are displayed in Fig.~\ref{fig1}. We only list the diagrams that contribute to the polarizabilities.

\begin{figure}[htbp]
    \centering
    \hspace*{\fill}
    \begin{subfigure}[b]{0.23\textwidth}
        \centering
        \includegraphics[page=1, width=1.0\textwidth]{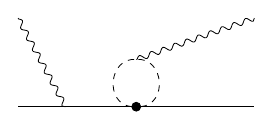}
        \caption*{($z_1$)}
    \end{subfigure}
    \hfill
    \begin{subfigure}[b]{0.23\textwidth}
        \centering
        \includegraphics[page=2, width=1.0\textwidth]{feynman2.pdf}
        \caption*{($z_2$)}
    \end{subfigure}
    \hfill
    \begin{subfigure}[b]{0.23\textwidth}
        \centering
        \includegraphics[page=4, width=1.0\textwidth]{feynman2.pdf}
        \caption*{($z_3$)}
    \end{subfigure}
    \hfill
    \begin{subfigure}[b]{0.23\textwidth}
        \centering
        \includegraphics[page=3, width=1.0\textwidth]{feynman2.pdf}
        \caption*{($z_4$)}
    \end{subfigure}
    \hfill
    \hspace*{\fill}
    \hspace*{\fill}
    \begin{subfigure}[b]{0.23\textwidth}
        \centering
        \includegraphics[page=5, width=1.0\textwidth]{feynman2.pdf}
        \caption*{($z_5$)}
    \end{subfigure}
    \hfill
        \begin{subfigure}[b]{0.23\textwidth}
        \centering
        \includegraphics[page=7, width=1.0\textwidth]{feynman2.pdf}
        \caption*{($z_6$)}
    \end{subfigure}
    \hfill
        \begin{subfigure}[b]{0.23\textwidth}
        \centering
        \includegraphics[page=6, width=1.0\textwidth]{feynman2.pdf}
        \caption*{($z_7$)}
    \end{subfigure}
    \hfill
    \begin{subfigure}[b]{0.23\textwidth}
    \centering
    \includegraphics[page=8, width=1.0\textwidth]{feynman2.pdf}
     \caption*{($z_8$)}
    \end{subfigure}
    \hfill
    \hspace*{\fill}
    \captionsetup{justification=raggedright, singlelinecheck=false}
    \caption{Some typical Feynman diagrams that do not contribute to the polarizabilities. The solid dot denotes a vertex or propagator from $\mathcal{L}^{(2)}$. Each diagram includes its corresponding crossed and symmetric diagrams.}
\label{fig2}
\end{figure}

There are some other Feynman diagrams at $\mathcal{O}(p^4)$ which exist but do not contribute to the polarizabilities, and we list some typical ones in Fig.~\ref{fig2}. We use the Coulomb gauge condition $v\cdot \epsilon = 0$ which causes that the first order vertices $\gamma \mathcal{B} \mathcal{B}$ and $\gamma \phi \phi \mathcal{B} \mathcal{B}$ vanish in the diagrams $z_1$, $z_2$, and $z_6$, and therefore the contributions from these diagrams are zero. For $z_3$  and $z_4$, the contraction of $v^\mu$ with $S_\mu$ appears in the result, making both of them vanish. Among $z_5$ and its crossed and symmetric diagrams, the corresponding vertices are identical and the only difference is a minus sign in the baryon propagator, and thus their contributions cancel each other out.

For $z_7$ in Fig.~\ref{fig2}, all vertices and propagators are independent of the photon momentum $k$ or $k'$, so the final result is a constant independent of the photon energy $\omega$. Its derivative with respect to $\omega$ is therefore zero, and hence it does not contribute to the polarizabilities. Furthermore, according to the Lehmann-Symanzik-Zimmermann formalism, the amplitudes of the external leg correction diagrams $z_8$ in Fig.~\ref{fig2} up to $\mathcal{O}(p^4)$ equal the $\mathcal{O}(p^2)$ contact term amplitudes multiplied by a renormalization constant $\sqrt{Z}$. Since the amplitudes from the $\mathcal{O}(p^2)$ contact term contributions do not vary with $\omega$, they do not contribute to the polarizabilities.

There is only one elastic part left in the diagram $(d_5+e_5)$ of Fig.~\ref{fig1}. The loop subdiagram of this reducible diagram is of order $\mathcal{O}(p^3)$, and thus it seems that a third order Lagrangian $\gamma \mathcal{B}\mathcal{B}$ is indeed needed to absorb its divergence. However, one does not need to introduce this kind of Lagrangian as discussed in Ref.~\cite{Bernard:1993ry} since the third order vertices enter only tree diagrams and therefore only contribute to the polynomial part of the amplitude, while the most general polynomial piece for $\gamma\gamma \mathcal{B}\mathcal{B}$ at $\mathcal{O}(p^4)$ is already given by the contact vertices with the LECs $a_i$. If we choose $a_1 = a_4 = \frac{9 g_1^2(2d_6+d_5)}{128\pi M_N F^2_0}L$, $a_2 = a_5 = -\frac{5 g_1^2 d_5}{192\pi M_N F^2_0}L$, $a_3 = a_6 = -\frac{g_1^2(12d_6-d_5)}{192\pi M_N F^2_0}L$, the divergences of the diagram $(d_5+e_5)$ can be precisely canceled.

\section{\label{sec:3}ANALYTICAL EXPRESSIONS}

This section presents the analytical expressions for the scalar and spin polarizabilities derived from the $\mathcal{O}(p^3)$ and $\mathcal{O}(p^4)$ diagrams. We systematically calculate the amplitude for each diagram, extract the coefficients $A_i$ following Eq.~(\ref{eq5}), and finally derive the analytical expressions following Eq.~(\ref{eq6}). The explicit expressions for the coefficients $A_i$ from each diagram are given in the Appendix.

\subsection{$\mathcal{O}(p^3)$ contributions to the polarizabilities}

For the scalar polarizabilities $\alpha_E$ and $\beta_M$, we have checked the consistency between our results for the general nonforward case and the forward Compton scattering in Ref.~\cite{Chen:2024xks}. The spin polarizabilities were not studied in Ref.~\cite{Chen:2024xks} and will be investigated in this work. We present the $\mathcal{O}(p^3)$ spin polarizabilities from diagrams $(a)-(g)$ in Fig.~1  of Ref.~\cite{Chen:2024xks} whose intermediate charmed baryons have spin 1/2,
\begin{equation}
\begin{aligned}
    &\gamma^{(a-g)}_{1,\xi} = x_\xi \frac{\alpha_{em} g^2_1}{48 \pi^2 M_\pi^2 F^2_\pi} + y_\xi \frac{\alpha_{em} g^2_1}{48 \pi^2 M_K^2 F^2_K}, \\
    & \gamma^{(a-g)}_{2,\xi} = \frac{1}{2} \gamma^{(a-g)}_{1,\xi}, \\
    & \gamma^{(a-g)}_{3,\xi} = \frac{1}{4} \gamma^{(a-g)}_{1,\xi}, \\
    & \gamma^{(a-g)}_{4,\xi} = - \frac{1}{4} \gamma^{(a-g)}_{1,\xi}.
\end{aligned}
\label{eq20}
\end{equation}
Here we define parameters $x_\xi$ and $y_\xi$ to simplify the expressions, with their specific values given in Table~\ref{tab1}.
\begin{table}
\caption{The values of parameters $x_\xi$, $y_\xi$, $m$ and $n$.}
\begin{ruledtabular}
\renewcommand{\arraystretch}{1.7}
\begin{tabular}{ccccccc}
 &$\Sigma_c^{++}$ & $\Sigma_c^{+}$ & $\Sigma_c^{0}$ & $\Xi_c^{'+}$ & $\Xi_c^{'0}$ & $\Omega_c^{0}$\\ \hline
 $x_\xi$ & $1$ & $2$ & $1$ & $\frac{1}{2}$ & $\frac{1}{2}$ & $0$ \\ 
 $y_\xi$ & $1$ & $\frac{1}{2}$ & $0$ & $2$ & $\frac{1}{2}$ & $1$ \\ 
 $m$ & $1$ & $2$ & $2$ & $3$ & $2$ & $\cdots$ \\ 
 $n$ & $1$ & $3$ & $\cdots$ & $2$ & $2$ & $2$ \\ 
\end{tabular}
\end{ruledtabular}
\label{tab1}
\end{table}
$\xi$ and $\chi$ denote the incoming baryon and meson, respectively. $\alpha_{em} = 1/137$ is the fine-structure constant. The $\mathcal{O}(p^3)$ spin polarizabilities from diagrams $(b')-(g')$ in Fig.~1 of Ref.~\cite{Chen:2024xks} whose intermediate charmed baryons have spin 3/2,
\begin{equation}
\begin{aligned}
    \gamma^{(b'-g')}_{1,\xi} =& x_\xi \frac{\alpha_{em} g_3^2 S^{(2)}_\pi}{288\pi^2 F_\pi^2 (M_\pi^2 - \delta^2)^3} \\
    & + y_\xi \frac{\alpha_{em} g_3^2 S^{(2)}_K}{288\pi^2 F_K^2 (M_K^2 - \delta^2)^3}, \\
    \gamma^{(b'-g')}_{2,\xi} =& \frac{\alpha_{em} C^2_{1,\xi}}{24M_N^2 \delta^2} + x_\xi \frac{\alpha_{em} g_3^2 S^{(3)}_\pi}{288\pi^2 F_\pi^2 (M_\pi^2 - \delta^2)^2} \\
    & + y_\xi \frac{\alpha_{em} g_3^2 S^{(3)}_K}{288\pi^2 F_K^2 (M_K^2 - \delta^2)^2}, \\
    \gamma^{(b'-g')}_{3,\xi} =& x_\xi \frac{\alpha_{em} g_3^2 S^{(3)}_\pi}{576\pi^2 F_\pi^2 (M_\pi^2 - \delta^2)^2} \\
    & + y_\xi \frac{\alpha_{em} g_3^2 S^{(3)}_K}{576\pi^2 F_K^2 (M_K^2 - \delta^2)^2}, \\
    \gamma^{(b'-g')}_{4,\xi} =& - \frac{\alpha_{em} C^2_{1,\xi}}{24M_N^2 \delta^2} - \gamma^{(b'-g')}_{3,\xi}
\end{aligned}
\label{eq21}
\end{equation}
For clarity, we define the common factor $S^{(2,3)}$ as
\begin{equation}
\begin{aligned}
    &S^{(2)}_\chi = -2 M_\chi^4 + \delta^4 + M_\chi^2 \delta (3R_\chi + \delta), \\
    &S^{(3)}_\chi = - M_\chi^2 + \delta (R_\chi + \delta),
\end{aligned}
\label{eq22}
\end{equation}
where $R_\chi = \sqrt{M^2_\chi - \delta^2} \arccos \frac{\delta}{M_\chi}$. The relations for $C_{1,\xi}$ and low-energy constants (LECs) are listed in Table~\ref{tab2}. 

\begin{table*}
\caption{The $C_{1,\xi}$, $C_{2,\xi}$, $C_{3,\xi}$, $\tilde{C}_{1,\xi,\chi}$,  $\tilde{C}_{2,\xi,\chi}$,  and  $\tilde{C}_{3,\xi,\chi}$ are expressed with the LECs.
}
\begin{ruledtabular}
\renewcommand{\arraystretch}{1.7}
\begin{tabular}{ccccccc}
 &$\Sigma_c^{++}$ & $\Sigma_c^{+}$ & $\Sigma_c^{0}$ & $\Xi_c^{'+}$ & $\Xi_c^{'0}$ & $\Omega_c^{0}$\\ \hline
 $C_{1,\xi}$ & $f_7+\frac{2}{3}f_6$ & $f_7+\frac{1}{6}f_6$ & $f_7-\frac{1}{3}f_6$ & $f_7+\frac{1}{6}f_6$ & $f_7-\frac{1}{3}f_6$ & $f_7-\frac{1}{3}f_6$ \\ 
 $C_{2,\xi}$ & $d_6+\frac{2}{3}d_5$ & $d_6+\frac{1}{6}d_5$ & $d_6-\frac{1}{3}d_5$ & $d_6+\frac{1}{6}d_5$ & $d_6-\frac{1}{3}d_5$ & $d_6-\frac{1}{3}d_5$ \\ 
 $C_{3,\xi}$ & $f_{10}+\frac{2}{3}f_9$ & $f_{10}+\frac{1}{6}f_9$ & $f_{10}-\frac{1}{3}f_9$ & $f_{10}+\frac{1}{6}f_9$ & $f_{10}-\frac{1}{3}f_9$ & $f_{10}-\frac{1}{3}f_9$ \\ \hline
 $\tilde{C}_{i,\xi,\pi}$ & $C_{i,\Sigma^+_c}$ & $\frac{1}{2}C_{i,\Sigma^{++}_c} + \frac{1}{2}C_{i,\Sigma^0_c}$ & $C_{i,\Sigma^+_c}$ & $C_{i,\Xi'^0_c}$ & $C_{i,\Xi'^+_c}$ & $\cdots$ \\
 $\tilde{C}_{i,\xi,K}$ & $C_{i,\Xi'^+_c}$ & $C_{i,\Xi'^0_c}$ & $\cdots$ & $\frac{1}{2}C_{i,\Sigma^{++}_c} + \frac{1}{2}C_{i,\Omega^0_c}$ & $C_{i,\Sigma^+_c}$ & $C_{i,\Xi'^+_c}$ \\
\end{tabular}
\end{ruledtabular}
\label{tab2}
\end{table*}

\subsection{$\mathcal{O}(p^4)$ contributions to the polarizabilities}

The $\mathcal{O}(p^4)$ results contain the divergence which can be removed with the modified minimal subtraction ($\widetilde{\text{MS}}$) scheme. The divergent $L$ is precisely canceled by the corresponding counterterm $a_2$, analogous to the case of nucleons \cite{Bernard:1993ry}.

The results for the $\mathcal{O}(p^4)$ electric polarizabilities from spin-1/2 contributions are
\begin{equation}
\begin{aligned}
    \alpha^{(c_2-g_3)}_{E,\xi} =& x_\xi \alpha^{(c_2-g_3)}_{E,m,\xi,\pi} + y_\xi \alpha^{(c_2-g_3)}_{E,n,\xi,K}, \\
    \alpha^{(h_2-h_5)}_{E,\xi} =& \frac{\alpha_{em}}{96 \pi^2}\sum_\chi \frac{1}{F^2_\chi}\left[c_{\xi,\chi,2}(3+6\ln \frac{M_\chi}{\mu}) - 2c_{\xi,\chi,1}\right].
\end{aligned}
\label{eq23}
\end{equation}
Here the subscripts $m$ and $n$ are given in Table~\ref{tab1}. $c_{\xi,\chi,i}$ is a linear combination of $c_{0-4}$, with the specific relation given in Eqs.~(\ref{eqc1.1}) and (\ref{eqc1.2}). The $\alpha^{(c_2-g_3)}_{E,n,\xi,\chi}$ are defined as
\begin{equation*}
\begin{aligned}
    \alpha^{(c_2-g_3)}_{E,1,\xi,\chi} =& \frac{\alpha_{em} g_1^2}{768\pi^2 F_\chi^2}\frac{187-16Q_\xi+238\ln \frac{M_\chi}{\mu}}{M_6}, \\
    \alpha^{(c_2-g_3)}_{E,2,\xi,\chi} =& \frac{\alpha_{em} g_1^2}{768\pi^2 F_\chi^2}\frac{91-16Q_\xi+46\ln \frac{M_\chi}{\mu}}{M_6}, \\
\end{aligned}
\end{equation*}
\begin{equation}
\begin{aligned}
    \alpha^{(c_2-g_3)}_{E,3,\xi,\chi} =& \frac{\alpha_{em} g_1^2}{768\pi^2 F_\chi^2}\frac{139-16Q_\xi+142\ln \frac{M_\chi}{\mu}}{M_6}.
\end{aligned}
\label{eq24}
\end{equation}

The results for the $\mathcal{O}(p^4)$ electric polarizabilities from spin-3/2 contributions are
\begin{equation}
\begin{aligned}
    \alpha^{(c_2'-g_3')}_{E,\xi} =& x_\xi \alpha^{(c_2'-g_3')}_{E,m,\xi,\pi} + y_\xi \alpha^{(c_2'-g_3')}_{E,n,\xi,K}.
\end{aligned}
\label{eq25}
\end{equation}
Here we define
\begin{equation}
\begin{aligned}
    \alpha^{(c_2'-g_3')}_{E,1,\xi,\chi} =& \frac{\alpha_{em}g^2_3}{1728 \pi^2 F^2_\chi(M_\chi^2-\delta^2)^2} \left(\frac{T^{(1)}_{\xi,\chi}}{M_{6^*}} + \frac{U^{(1)}_{\chi}}{M_6}\right), \\
    \alpha^{(c_2'-g_3')}_{E,2,\xi,\chi} =& \frac{\alpha_{em}g^2_3}{1728 \pi^2 F^2_\chi(M_\chi^2-\delta^2)^2} \left(\frac{T^{(1)}_{\xi,\chi}}{M_{6^*}} + \frac{U^{(2)}_{\chi}}{M_6}\right), \\
    \alpha^{(c_2'-g_3')}_{E,3,\xi,\chi} =& \frac{\alpha_{em}g^2_3}{1728 \pi^2 F^2_\chi(M_\chi^2-\delta^2)^2} \left(\frac{T^{(1)}_{\xi,\chi}}{M_{6^*}} + \frac{U^{(3)}_{\chi}}{M_6}\right),
\end{aligned}
\label{eq26}
\end{equation}
and
\begin{equation}
\begin{aligned}
    T^{(1)}_{\xi,\chi} =& M^2_\chi \delta\left[(240-12Q_\xi)R_\chi - (505-60Q_\xi)\delta\right] \\
    & + \delta^3\left[-147R_\chi + (299-36Q_\xi)\delta\right] \\
    & + (206-24Q_\xi)M^4_\chi + 147(M^2_\chi - \delta^2)^2 \ln \frac{M_\chi}{\mu}, \\
    U^{(1)}_{\chi} =& M^2_\chi \delta\left(336R_\chi - 414\delta\right) + \delta^3\left(-198R_\chi + 276\delta\right) \\
    & + 138M^4_\chi + 198(M^2_\chi - \delta^2)^2\ln \frac{M_\chi}{\mu}, \\
    U^{(2)}_{\chi} =& M^2_\chi \delta\left(-144R_\chi + 162\delta\right) + \delta^3\left(90R_\chi - 108\delta\right) \\
    & - 54M^4_\chi - 90(M^2_\chi - \delta^2)^2\ln \frac{M_\chi}{\mu}, \\
    U^{(3)}_{\chi} =& M^2_\chi \delta\left(96R_\chi - 126\delta\right) + \delta^3\left(-54R_\chi + 84\delta\right) \\
    & + 42M^4_\chi + 54(M^2_\chi - \delta^2)^2\ln \frac{M_\chi}{\mu}.
\end{aligned}
\label{eq27}
\end{equation}

The results for the $\mathcal{O}(p^4)$ magnetic polarizabilities from both the spin-1/2 and 3/2 contributions are
\begin{equation}
\begin{aligned}
    \beta^{(c_2^{(\prime)}-g_3^{(\prime)})}_{M,\xi} =& x_\xi \beta_{M,\xi,\pi} + y_\xi \beta_{M,\xi,K}, \\
    \beta^{(h_2-h_5)}_{M,\xi} =& \frac{\alpha_{em}}{24 \pi^2}\sum_\chi \frac{1}{F^2_\chi}\left[-\frac{c_{\xi,\chi,2}}{4}(1+2\ln \frac{M_\chi}{\mu}) + \frac{c_{\xi,\chi,1}}{2}\right].
\end{aligned}
\label{eq28}
\end{equation}
where
\begin{equation}
\begin{aligned}
    \beta^{(c_2-g_3)}_{M,\xi,\chi} =& \frac{\alpha_{em} g_1^2}{768\pi^2 F_\chi^2}\left[\frac{51+94\ln \frac{M_\chi}{\mu}}{M_6} -(1+2\ln \frac{M_\chi}{\mu}) \right. \\
    &\left. \frac{24(C_{2,\xi}+\tilde{C}_{2,\xi,\chi})}{M_N}\right], \\
    \beta^{(c_2'-g_3')}_{M,\xi,\chi} =& \frac{\alpha_{em}g_3}{1728 \pi^2 F^2_\chi \delta(M_\chi^2-\delta^2)} \\
    & \left(\frac{T^{(2)}_\chi}{M_{6^*}} + \frac{U^{(4)}_{\chi}}{M_6} + \frac{V^{(1)}_{\xi,\chi}}{M_N}\right),\\
\end{aligned}
\label{eq29}
\end{equation}
and
\begin{equation}
\begin{aligned}
    T^{(2)}_\chi =& g_3 \left[98M_\chi^2 \delta + 147 \delta^2 R_\chi - 98 \delta^3 \right.\\
    &\left. + 147 \delta (M_\chi^2 - \delta^2)\ln \frac{M_\chi}{\mu}\right], \\
    U^{(4)}_{\chi} =& g_3\left[-2M_\chi^2 \delta + 6 \delta^2 R_\chi + 2\delta^3 \right.\\
    &\left. + 6\delta (M_\chi^2 - \delta^2)\ln \frac{M_\chi}{\mu}\right], \\
    V^{(1)}_{\xi,\chi} =& g_1 \left[\tilde{C}_{1,\xi,\chi}\left(-18 \pi M_\chi^3 + 36M_\chi^2 R_\chi + 30M_\chi^2 \delta \right.\right.\\
    &\left.+ 18\pi M_\chi \delta^2  - 36\delta^2 R_\chi - 30\delta^3 \right)+18 C_{1,\xi} \\
    &\left.(M_\chi^2 - \delta^2)\pi M_\chi\right] + g_3 \left[C_{2,\xi}\left(6 M_\chi^2 \delta + 36 \delta^2 R_\chi \right.\right.\\
    &\left. - 6\delta^3\right) + \tilde{C}_{3,\xi,\chi} \left(10 M_\chi^2 \delta \left. - 60 \delta^2 R_\chi - 10\delta^3 \right)\right] \\
    & + 10 g_5 C_{1,\xi} (M_\chi^2 - \delta^2)(6R_\chi + 7\delta) + 3 \\
    & \left[4 g_3 (3 C_{2,\xi} - 5\tilde{C}_{3,\xi,\chi}) - 4(3 g_1 \tilde{C}_{1,\xi,\chi} + 5 g_5 C_{1,\xi})\right] \\
    &\delta (M_\chi^2 - \delta^2)\ln \frac{M_\chi}{\mu}.
\end{aligned}
\label{eq30}
\end{equation}
The expressions for $C_{1,\xi}$, $C_{2,\xi}$, $C_{3,\xi}$, $\tilde{C}_{1,\xi,\chi}$,  $\tilde{C}_{2,\xi,\chi}$, and $\tilde{C}_{3,\xi,\chi}$ are summarized in Table~\ref{tab2}. 

The results for the $\mathcal{O}(p^4)$ spin polarizabilities $\gamma_1$, $\gamma_2$, $\gamma_3$, and $\gamma_4$ are
\begin{equation}
\begin{aligned}
    \gamma_{i,\xi} =& x_\xi \gamma_{i,m,\xi,\pi} + y_\xi \gamma_{i,n,\xi,K}, \qquad &i=1, \\
    \gamma_{i,\xi} =& x_\xi \gamma_{i,\xi,\pi} + y_\xi \gamma_{i,\xi,K}, \qquad &i=2,3,4,
\end{aligned}
\label{eq31}
\end{equation}
where we define
\begin{equation*}
\begin{aligned}
    \gamma^{(c_2-g_3)}_{1,1,\xi,\chi} =& -\frac{\alpha_{em} g_1^2}{384\pi M_\chi F_\chi^2}\frac{29+6Q_\xi}{M_6}, \\
    \gamma^{(c_2-g_3)}_{1,2,\xi,\chi} =& -\frac{\alpha_{em} g_1^2}{384\pi M_\chi F_\chi^2}\frac{-3+6Q_\xi}{M_6}, \\
\end{aligned}
\end{equation*}
\begin{equation}
\begin{aligned}
    \gamma^{(c_2-g_3)}_{1,3,\xi,\chi} =& -\frac{\alpha_{em} g_1^2}{384\pi M_\chi F_\chi^2}\frac{13+6Q_\xi}{M_6}, \\
    \gamma^{(c_2'-g_3')}_{1,1,\xi,\chi} =& \frac{\alpha_{em} g^2_3}{576\pi^2 F^2_\chi(M_\chi^2 - \delta^2)^3}\left(\frac{T^{(3)}_{\xi,\chi}}{M_{6^*}} + \frac{U^{(5)}_\chi}{M_6}\right), \\
    \gamma^{(c_2'-g_3')}_{1,2,\xi,\chi} =& \frac{\alpha_{em} g^2_3}{576\pi^2 F^2_\chi(M_\chi^2 - \delta^2)^3}\left(\frac{T^{(3)}_{\xi,\chi}}{M_{6^*}} + \frac{U^{(6)}_\chi}{M_6}\right), \\
    \gamma^{(c_2'-g_3')}_{1,3,\xi,\chi} =& \frac{\alpha_{em} g^2_3}{576\pi^2 F^2_\chi(M_\chi^2 - \delta^2)^3}\left(\frac{T^{(3)}_{\xi,\chi}}{M_{6^*}} + \frac{U^{(7)}_\chi}{M_6}\right),\\
    \gamma^{(c_2-g_3)}_{2,\xi,\chi} =& \frac{\alpha_{em} g_1^2}{384\pi M_\chi F_\chi^2}\left[-\frac{7}{M_6} + \frac{2(2C_{2,\xi} + \tilde{C}_{2,\xi,\chi})}{M_N}\right], \\
    \gamma^{(c_2'-g_3')}_{2,\xi,\chi} =& \frac{\alpha_{em} g_3}{576\pi^2 F^2_\chi(M^2_\chi - \delta^2)^2}\left(\frac{7T^{(4)}_\chi}{M_{6^*}} + \frac{V^{(2)}_{\xi,\chi}}{M_N}\right),\\
    \gamma^{(c_2-g_3)}_{3,\xi,\chi} =& \frac{\alpha_{em} g_1^2}{768\pi M_\chi F_\chi^2}\frac{-3+2Q_\xi}{M_6}, \\
    \gamma^{(c_2'-g_3')}_{3,\xi,\chi} =& \frac{\alpha_{em} g_3}{1152\pi^2 F^2_\chi (M^2_\chi - \delta^2)^2} \\
    & \left(\frac{(3+2Q_\xi)T^{(4)}_\chi}{M_{6^*}} - \frac{4Q_\xi T^{(4)}_\chi}{M_6}\right), \\
    \gamma^{(c_2-g_3)}_{4,\xi,\chi} =& \frac{\alpha_{em} g_1^2}{768\pi M_\chi F_\chi^2}\left(\frac{11}{M_6} - \frac{8C_{2,\xi} + 4\tilde{C}_{2,\xi,\chi}}{M_N}\right), \\
    \gamma^{(c_2'-g_3')}_{4,\xi,\chi} =& \frac{\alpha_{em} g_3}{3456\pi^2 F^2_\chi (M^2_\chi - \delta^2)^2} \left(\frac{-33T^{(4)}_\chi}{M_{6^*}} + \frac{V^{(3)}_{\xi,\chi}}{M_N}\right),
\end{aligned}
\label{eq32}
\end{equation}
and
\begin{equation*}
\begin{aligned}
    T^{(3)}_{\xi,\chi} =& M^4_\chi((5+6Q_\xi)R_\chi - (7+10Q_\xi)\delta) - 4(1+Q_\xi)\delta^5 \\
    & + M^2_\chi \delta^2(-2R_\chi + (11+14Q_\xi)\delta), \\
    T^{(4)}_\chi =& g_3 \left[M^2_\chi(R_\chi - \delta) + \delta^3\right], \\
    U^{(5)}_\chi =& 4 M^4_\chi(6R_\chi - 11\delta) - 14\delta^5 + 2M^2_\chi \delta^2(3R_\chi + 29\delta), \\
    U^{(6)}_\chi =& 2 M^4_\chi(-4R_\chi + 6\delta) + 6 \delta^5 + 2M^2_\chi \delta^2(R_\chi - 9\delta), \\
    U^{(7)}_\chi =& 8 M^4_\chi(R_\chi - 2\delta) - 4 \delta^5 + 4M^2_\chi \delta^2(R_\chi + 5\delta), \\
    V^{(2)}_{\xi,\chi} =& - \frac{2}{3}g_3 (6C_{2,\xi} + \tilde{C}_{3,\xi,\chi}) \left[M^2_\chi(R_\chi - \delta) + \delta^3\right]\\
    & + C_{1,\xi}\left(\frac{M^2_\chi}{\delta^2} - 1\right) \left\{-3g_1 \pi M_\chi (M^2_\chi - \delta^2) + 2g_5\right.\\
    &\left.\left[-M^2_\chi (5R_\chi + \frac{16}{3}\delta) + 2\delta^2(R_\chi + \frac{8}{3}\delta)\right]\right\} \\
    & + \tilde{C}_{1,\xi,\chi} g_1 \left(\frac{M^2_\chi}{\delta^2} - 1\right)\left[9 \pi M_\chi (M^2_\chi - \delta^2)\right.\\
    &\left. - 18 M^2_\chi (R_\chi + \delta)  + 2\delta^2(4R_\chi + 9\delta)\right] \\
    & + 4 g_5 C_{1,\xi} \frac{1}{\delta} (M^2_\chi - \delta^2)^2 \ln \frac{M_\chi}{\mu}, \\
\end{aligned}
\end{equation*}
\begin{equation}
\begin{aligned}
    V^{(3)}_{\xi,\chi} =& \frac{1}{\delta^2}\left\{3 g_1\left[C_{1,\xi}(M_\chi^2-\delta^2)^2(6\pi M_\chi + \delta)-18 \tilde{C}_{1,\xi,\chi} \right.\right.\\
    & (M_\chi^2 - \delta^2)\left(\pi M_\chi^3 - \pi M_\chi \delta^2 - 2M_\chi^2 (R_\chi + \delta)\right. \\
    &\left.\left. + \delta^2(R_\chi + 2\delta)\right)\right] + 24 C_{2,\xi} \delta^2 T^{(4)}_\chi + g_5 C_{1,\xi} \\
    & (M_\chi^2 - \delta^2)\left(-5\delta^2(6R_\chi + 13\delta)\right. \\
    &\left. +5M_\chi^2(12R_\chi + 13\delta)\right) + 6C_{1,\xi}(3g_1 - 5g_5) \\
    & \left.\delta (M_\chi^2 - \delta^2)^2 \ln \frac{M_\chi}{\mu}\right\}.
\end{aligned}
\label{eq33}
\end{equation}

\section{\label{sec:4}NUMERICAL RESULTS AND Discussion}

The analytical expressions above contain 21 undetermined LECs: the axial couplings $g_{1,2,3,5}$ from the LO Lagrangian $\mathcal{L}^{(1)}_{\mathcal{B}\phi}$ in Eq.~(\ref{eq16}), the constants $f_{6,7,9,10}$ and $d_{5,6}$ from the NLO Lagrangian $\mathcal{L}^{(2)}_{\mathcal{B}\phi}$ in Eq.~(\ref{eq17}), $c_{0\text{-}4}$ from $\mathcal{L}^{(2)}_{\mathcal{B}\phi \phi}$, and $a_{1\text{-}6}$ from $\mathcal{L}^{(4)}_{\mathcal{B}\phi}$. The axial coupling constants $g_{1,2,3,5}$ can be related to the quark model and determined by the experimental decay widths of charmed baryons \cite{Chen:2024xks}: 
\begin{equation}
\begin{aligned}
    &g_1 = 0.867 \pm 0.087 \pm 0.035, \\
    &g_2 = -\sqrt{\frac{3}{8}} g_1 = -0.531 \pm 0.053 \pm 0.021, \\
    &g_3 = \frac{\sqrt{3}}{2} g_1 = 0.751 \pm 0.076 \pm 0.030, \\
    &g_5 = - \frac{3}{2} g_1 = - 1.301 \pm 0.131 \pm 0.053.
\end{aligned}
\label{eq34}
\end{equation}
The LECs $f_{6,7,9,10}$ and $d_{5,6}$ appear not individually but only in the specific combinations $C_{1,\xi}$, $C_{2,\xi}$, and $C_{3,\xi}$. 
These coefficients were determined by the magnetic moments \cite{Wang:2018gpl}, 
\begin{equation}
\begin{aligned}
    C_{1,\xi} = -\sqrt{6} \frac{\mu_{\xi^* \rightarrow \xi + \gamma}}{\mu_N};\qquad 
    C_{2,\xi} = \frac{\mu_\xi}{\mu_N}; \qquad 
    C_{3,\xi} = \frac{\mu_{\xi^*}}{\mu_N}.
\end{aligned}
\label{eq35}
\end{equation}
The magnetic moments of these charmed hadrons can also be calculated within the quark model and the values from Refs.~\cite{Chen:2024xks, Wang:2018gpl} are used here. The coupling constants $c_i$ were determined by the SU(4) symmetry in Ref. \cite{Liu:2012uw},
\begin{equation}
\begin{aligned}
    c_0 =& -0.61 \pm 0.10 \enspace\text{GeV}^{-1}, \\
    c_1 =& -0.98 \pm 0.01 \enspace\text{GeV}^{-1}, \\
    c_2 =& -2.07 \pm 1.87 \enspace\text{GeV}^{-1} - 2\frac{\alpha'}{4\pi F_0}, \\
    c_3 =& -0.84 \pm 0.21 \enspace\text{GeV}^{-1}, \\
    c_4 =& \frac{\alpha'}{4\pi F_0}, \\
\end{aligned}
\end{equation}
where the dimensionless LEC $\alpha'$ varied in the range of $[-1,1]$ to estimate the relevant errors \cite{Liu:2012uw}. The Lagrangian $\mathcal{L}^{(4)}_{\mathcal{B}\phi}$ only contributes to the electromagnetic polarizabilities rather than spin ones, and currently we approximately neglect it as in the study of nucleon electric polarizability \cite{Bernard:1993ry}.

\begin{table*}
\captionsetup{justification=raggedright, singlelinecheck=false}
\caption{The $\mathcal{O}(p^3)$ and $\mathcal{O}(p^3+p^4)$ numerical results for electromagnetic polarizabilities of the spin-$1/2$ singly charmed baryons in units of $10^{-4}$ $\text{fm}^3$. In all results, we use the root-sum-square method to calculate the uncertainties.}
\begin{ruledtabular}
\renewcommand{\arraystretch}{1.7}
\begin{tabular}{ccccccc}
 &$\Sigma_c^{++}$ & $\Sigma_c^{+}$ & $\Sigma_c^{0}$ & $\Xi_c^{'+}$ & $\Xi_c^{'0}$ & $\Omega_c^{0}$\\ \hline
 $\alpha^{\mathcal{O}(p^3)}_E$ & 4.70(81) & 8.22(142) & 3.91(68) & 3.54(60) & 2.35(40) & 0.79(13) \\ 
 $\alpha^{\mathcal{O}(p^3+p^4)}_E$ & 5.12(90) & 9.63(154) & 4.88(89) & 4.34(69) & 3.02(53) & 1.16(17) \\ \hline
 $\beta^{\mathcal{O}(p^3)}_M$ & 3.71(66) & 0.88(15) & 2.75(30) & 0.42(7) & 2.33(18) & 1.92(15) \\ 
 $\beta^{\mathcal{O}(p^3+p^4)}_M$ & 7.74(148) & 0.44(51) & 0.28(36) & 0.48(44) & 0.09(36) & -0.10(41) \\ 
\end{tabular}
\end{ruledtabular}
\label{tab3}
\end{table*}

The corresponding numerical results for the electromagnetic polarizabilities of the spin-$1/2$ singly charmed baryons are listed in Table~\ref{tab3} with the mass scale $\mu = 1 \enspace\text{GeV}$. The contributions at order $\mathcal{O}(p^4)$ are generally small for the electric polarizability, not exceeding $50\%$ of those at order $\mathcal{O}(p^3)$, which exhibits good convergence. The pions contribute to $\alpha_E$ more dominantly than kaons, and thus $\alpha_E|_{\Sigma^{++}_c}$ and $\alpha_E|_{\Sigma^{+}_c}$ are relatively larger while $\alpha_E|_{\Omega^0_c}$ is relatively smaller. The electric polarizabilities of all singly charmed baryons in the sextet are smaller than those of nucleons, e.g. $\alpha_E|_{\Sigma^{+}_c}\approx 4/5\, \alpha_E|_{p}$, which means that they are less affected than nucleons in a static electric field.

For magnetic polarizabilities, the $\mathcal{O}(p^4)$ corrections are roughly comparable to the $\mathcal{O}(p^3)$ contributions overall, similar to the nucleon case \cite{Holstein:2013kia}. Now we focus on the corrections at fourth order from the $f_i$ LECs. The large fourth order contributions mainly appear in the three polarizabilities $\beta_M$, $\gamma_2$ and $\gamma_4$ related to the magnetic field. We take the $\beta_M$ of $\Sigma^{++}_c$ as an example. For the spin-3/2 contributions, the $\mathcal{O}(p^4)$ contribution related to $f_i$ is $\beta_M = 4.34 \times 10^{-4} \text{fm}^3$, and that unrelated to $f_i$ is $\beta_M = -0.09 \times 10^{-4} \text{fm}^3$. It can be seen that the contributions from the terms involving $f_i$ are indeed larger. However, this enhancement is not actually caused by $f_i$ itself, and the real reason lies in the additional large factor $\sum_\chi M_\chi/\delta$ associated with the magnetic transition subprocess. There is no such factor $\sum_\chi M_\chi/\delta$ for the spin-1/2 contributions, and therefore the $\mathcal{O}(p^4)$ contribution related to $f_i$ ($\beta_M = 0.90 \times 10^{-4} \text{fm}^3$) is of the same order as that unrelated to $f_i$ ($\beta_M = -1.12 \times 10^{-4} \text{fm}^3$). Moreover, because of the difference in the magnitude of the transition magnetic moment, the $\mathcal{O}(p^4)$ contributions of $\Sigma^+_c$ and $\Xi^{'+}_c$ are relatively smaller, while the $\mathcal{O}(p^4)$ contributions of the other four particles are significantly larger.

For the proton electric polarizability $\alpha_E$ in HBChPT \cite{Bernard:1995dp, Bernard:1993bg}, the correction from $\mathcal{O}(p^3)$ to $\mathcal{O}(p^4)$ is indeed relatively moderate. The situation is entirely different for the magnetic polarizability $\beta_M$. In HBChPT, $\beta_M$ increases from $1.22$ at $\mathcal{O}(p^3)$ to $3.5$ for the proton at $\mathcal{O}(p^4)$ \cite{Bernard:1995dp, Bernard:1993bg}, corresponding to an $\mathcal{O}(p^4)$ correction of nearly $200\%$. Our results likewise exhibit substantial $\mathcal{O}(p^4)$ contributions similar to those found in the proton sector. 

The numerical results for the spin polarizabilities of the spin-$1/2$ singly charmed baryons are reported in Table~\ref{tab4}. The quantities $\gamma_0$ and $\gamma_\pi$, which are related to forward and backward Compton scattering, are linear combinations of the four spin polarizabilities and correspond directly to experimentally measurable physical quantities \cite{Holstein:2013kia}
\begin{equation}
\begin{aligned}
    \gamma_0 &= - \gamma_{E1E1} - \gamma_{M1M1} - \gamma_{E1M2} - \gamma_{M1E2}, \\
    \gamma_\pi &= - \gamma_{E1E1} + \gamma_{M1M1} - \gamma_{E1M2} + \gamma_{M1E2}.
\end{aligned}
\label{eq36}
\end{equation}
We therefore provide their values in Table~\ref{tab4} for comparison.

\begin{table*}
\captionsetup{justification=raggedright, singlelinecheck=false}
\caption{The $\mathcal{O}(p^3)$ and $\mathcal{O}(p^3+p^4)$ numerical results for spin polarizabilities of the spin-$1/2$ singly charmed baryons in units of $10^{-4}$ $\text{fm}^4$.}
\begin{ruledtabular}
\renewcommand{\arraystretch}{1.7}
\begin{tabular}{ccccccc}
 &$\Sigma_c^{++}$ & $\Sigma_c^{+}$ & $\Sigma_c^{0}$ & $\Xi_c^{'+}$ & $\Xi_c^{'0}$ & $\Omega_c^{0}$\\ \hline
 $\gamma^{\mathcal{O}(p^3)}_{E1E1}$ & -1.28(31) & -2.48(55) & -1.22(29) & -0.72(23) & -0.64(16) & -0.06(2) \\ 
 $\gamma^{\mathcal{O}(p^3+p^4)}_{E1E1}$ & -0.32(7) & -2.29(55) & -1.25(30) & -0.50(12) & -0.66(16) & -0.06(2) \\ \hline
 $\gamma^{\mathcal{O}(p^3)}_{M1M1}$ & -4.98(97) & -0.50(13) & -3.67(43) & -0.21(7) & -3.19(25) & -2.72(23) \\ 
 $\gamma^{\mathcal{O}(p^3+p^4)}_{M1M1}$ & 0.72(130) & 0.02(34) & -5.47(65) & 0.85(44) & -5.45(59) & -5.39(67) \\ \hline
 $\gamma^{\mathcal{O}(p^3)}_{E1M2}$ & 0.25(6) & 0.48(12) & 0.24(6) & 0.14(4) & 0.12(3) & 0.01(0) \\ 
 $\gamma^{\mathcal{O}(p^3+p^4)}_{E1M2}$ & 0.26(7) & 0.46(12) & 0.21(5) & 0.13(4) & 0.11(3) & 0.01(0) \\ \hline
 $\gamma^{\mathcal{O}(p^3)}_{M1E2}$ & 0.25(6) & 0.48(12) & 0.24(6) & 0.14(4) & 0.12(3) & 0.01(0) \\ 
 $\gamma^{\mathcal{O}(p^3+p^4)}_{M1E2}$ & 0.66(11) & 0.47(12) & -0.11(6) & 0.15(4) & -0.08(3) & -0.06(1) \\ \hline
 $\gamma^{\mathcal{O}(p^3)}_0$ & 5.76(100) & 2.02(48) & 4.42(49) & 0.65(16) & 3.58(28) & 2.76(23) \\
 $\gamma^{\mathcal{O}(p^3+p^4)}_0$ & -1.33(129) & 1.35(59) & 6.62(70) & -0.63(47) & 6.08(59) & 5.51(67) \\
 $\gamma^{\mathcal{O}(p^3)}_\pi$ & -3.70(100) & 1.96(48) & -2.45(49) & 0.51(16) & -2.55(28) & -2.67(23) \\ 
 $\gamma^{\mathcal{O}(p^3+p^4)}_\pi$ & 1.44(130) & 2.32(55) & -4.54(78) & 1.37(45) & -4.98(66) & -5.39(69) \\ 
\end{tabular}
\end{ruledtabular}
\label{tab4}
\end{table*}

Table~\ref{tab4} shows that the corrections from $\mathcal{O}(p^4)$ to the polarizabilities $\gamma_{E1E1}$ and $\gamma_{E1M2}$, which represent the electric field response, are relatively smaller, while the corrections to the magnetic field response are significantly larger and the reason is the same as that discussed earlier for the magnetic polarizability. This shows the same trend as the electromagnetic polarizabilities. Among these four spin polarizabilities, the magnetic dipole polarizability in the magnetic field gradient is the strongest. In general, we find that the spin polarizabilities of most singly charmed baryons are smaller than those of the proton. This is expected since the large mass of singly charmed baryons generally suppresses polarization.

\begin{table*}
\captionsetup{justification=raggedright, singlelinecheck=false}
\caption{Comparison between the contributions from the spin-$1/2$ and $3/2$ intermediate states for the $\mathcal{O}(p^4)$ electromagnetic and spin polarizabilities in units of $10^{-4}$ $\text{fm}^3$ and $10^{-4}$ $\text{fm}^4$, respectively.}
\begin{ruledtabular}
\renewcommand{\arraystretch}{1.7}
\begin{tabular}{cccccccc}
 & spin & $\Sigma_c^{++}$ & $\Sigma_c^{+}$ & $\Sigma_c^{0}$ & $\Xi_c^{'+}$ & $\Xi_c^{'0}$ & $\Omega_c^{0}$\\ \hline
 \multirow{2}{*}{$\alpha_{E}$} & 1/2 & 0.40(65) & 1.27(61) & 0.90(58) & 0.68(35) & 0.60(32) & 0.28(6) \\ 
 & 3/2 & 0.01(0) & 0.14(3) & 0.08(2) & 0.12(3) & 0.07(2) & 0.09(1) \\ \hline
 \multirow{2}{*}{$\beta_{M}$} & 1/2 & -0.22(26) & -0.83(25) & -1.04(24) & -0.72(14) & -0.59(13) & -0.20(5) \\ 
 & 3/2 & 4.25(97) & 0.39(29) & -1.43(29) & 0.78(35) & -1.65(35) & -1.82(41) \\ \hline
 \multirow{2}{*}{$\gamma_{E1E1}$} & 1/2 & 1.11(24) & 0.20(4) & -0.03(1) & 0.25(5) & -0.02(0) & -0.01(0) \\ 
 & 3/2 & -0.15(3) & -0.03(1) & 0.00(0) & -0.03(1) & 0.00(0) & 0.00(0) \\ \hline
 \multirow{2}{*}{$\gamma_{M1M1}$} & 1/2 & -0.30(8) & 0.05(5) & 0.29(7) & 0.07(2) & 0.16(4) & 0.05(1) \\ 
 & 3/2 & 6.00(134) & 0.47(40) & -2.08(39) & 1.00(51) & -2.42(49) & -2.71(59) \\ \hline
 \multirow{2}{*}{$\gamma_{E1M2}$} & 1/2 & 0.01(0) & -0.02(1) & -0.03(1) & -0.01(1) & -0.02(0) & -0.01(0) \\ 
 & 3/2 & 0.00(0) & 0.00(0) & 0.01(0) & 0.00(0) & 0.00(0) & 0.00(0) \\ \hline
 \multirow{2}{*}{$\gamma_{M1E2}$} & 1/2 & -0.04(1) & -0.07(2) & -0.03(1) & -0.03(1) & -0.02(0) & -0.01(0) \\ 
 & 3/2 & 0.46(8) & 0.06(5) & -0.31(5) & 0.04(2) & -0.18(3) & -0.06(1) \\
\end{tabular}
\end{ruledtabular}
\label{tab5}
\end{table*}

In Table~\ref{tab5}, we compare the contributions from spin-1/2 and spin-3/2 intermediate states at $\mathcal{O}(p^4)$. For $\alpha_E$, $\gamma_{E1E1}$ and $\gamma_{E1M2}$, the spin-3/2 contributions are generally significantly smaller than those from spin-1/2 states, though still non-negligible. By contrast, for $\beta_M$, $\gamma_{M1M1}$ and $\gamma_{M1E2}$, the spin-3/2 contributions are larger, indicating that excited states dominate these three polarizabilities in singly charmed baryons. The relative size of the spin-3/2 contributions we computed is consistent with the relative magnitude of the $\Delta$ contributions found in earlier nucleon calculations, where the $\Delta$ contribution is also relatively smaller for $\alpha_E$, $\gamma_{E1E1}$ and $\gamma_{E1M2}$ of nucleon but exceeds those from the nucleon for $\beta_M$, $\gamma_{M1M1}$ and $\gamma_{M1E2}$ \cite{Gellas:2000mx, Hemmert:1996xg}. 

The procedure for the polarizabilities of singly bottom baryons is virtually identical to that for the singly charmed baryons except that the parameters must be substituted. The coupling constants $g_{1,2,3,5}$ and the mass difference $\delta$ are replaced with \cite{Chen:2024xks}:
\begin{equation}
\begin{aligned}
    &g_{1,b} = 1.144 \pm 0.115 \pm 0.040, \\
    &g_{2,b} = -0.701 \pm 0.070 \pm 0.024, \\
    &g_{3,b} = 0.991 \pm 0.100 \pm 0.035, \\
    &g_{5,b} = -1.716 \pm 0.173 \pm 0.060, \\
    &\delta_{b} = 20 \enspace\text{MeV}.
\end{aligned}
\label{eq37}
\end{equation}
The magnetic moments in Eq.~(\ref{eq35}) for the $C_{1,\xi}$, $C_{2,\xi}$, $C_{3,\xi}$ should be changed to those of bottom baryons which can be found in Ref.~\cite{Chen:2024xks}.  We approximately use the same values of $c_i$ as in the single charm case according to the heavy quark symmetry.

\begin{table*}
\captionsetup{justification=raggedright, singlelinecheck=false}
\caption{The $\mathcal{O}(p^3+p^4)$ numerical electromagnetic and spin polarizabilities of the spin-$1/2$ singly bottom baryons in units of $10^{-4}$ $\text{fm}^3$ and $10^{-4}$ $\text{fm}^4$, respectively.}
\begin{ruledtabular}
\renewcommand{\arraystretch}{1.7}
\begin{tabular}{ccccccc}
 &$\Sigma_b^{+}$ & $\Sigma_b^{0}$ & $\Sigma_b^{-}$ & $\Xi_b^{'0}$ & $\Xi_b^{'-}$ & $\Omega_b^{-}$\\ \hline
 $\alpha_E$ & 9.66(150) & 16.85(259) & 7.95(128) & 7.47(114) & 4.84(76) & 1.71(25) \\ 
 $\beta_M$ & 45.95(798) & 10.43(291) & -0.80(86) & 14.46(410) & -2.53(110) & -3.98(137) \\ 
 $\gamma_{E1E1}$ & -1.63(43) & -3.81(101) & -1.73(46) & -1.16(31) & -0.89(24) & -0.06(2) \\ 
 $\gamma_{M1M1}$ & 38.40(2746) & 28.06(890) & -35.23(792) & 52.23(1553) & -39.43(853) & -42.27(988) \\ 
 $\gamma_{E1M2}$ & 0.41(11) & 0.75(20) & 0.36(10) & 0.26(7) & 0.19(5) & 0.02(0) \\
 $\gamma_{M1E2}$ & 4.07(63) & 2.55(54) & -0.93(24) & 1.12(23) & -0.53(13) & -0.23(5) \\ 
\end{tabular}
\end{ruledtabular}
\label{tab6}
\end{table*}

In Table~\ref{tab6}, we present the $\mathcal{O}(p^3)+\mathcal{O}(p^4)$ results for the polarizabilities of singly bottom baryons. $\beta_M$ for $\Sigma^+_b$ is $45.95=18.44|_{\mathcal{O}(p^3)}+27.51|_{\mathcal{O}(p^4)}$, considerably larger than those of other baryons. This feature was already present in its $\mathcal{O}(p^3)$ calculation in Ref.~\cite{Chen:2024xks} and the reason is same as we discuss in the charmed case. The values of $\gamma_{M1M1}$ are significantly larger than those of singly charmed baryons. This enhancement can also be traced to the dominance of the $b'$, $(d''_5+e''_5)$ and $(d'''_5+e'''_5)$ diagrams up to $\mathcal{O}(p^4)$ where there are $1/\delta^2$ factors as the mass splitting $\delta$ in singly bottom baryons is smaller than in singly charmed baryons.

We also estimate the effects of $\mathcal{O}(p^4)$ LECs alternatively by using the nonanalytic-dominance approximation adopted in Refs.~\cite{Bijnens:1995yn, Wang:2019mhm} which assume that the NLO contact terms largely cancel the analytic part of the NLO loop contributions. We compare the nonanalytic-dominance results with those using the zero $\mathcal{O}(p^4)$ LEC assumption in Table~\ref{tab10}. The difference between them provides a practical measure of the sensitivity to the $\mathcal{O}(p^4)$ LECs. For the great majority of the polarizabilities considered, the relative differences between the two scenarios are below 15\%. Some polarizabilities of few baryons change much with these two different scenarios and deserve further investigation in future.

\begin{table}
\captionsetup{justification=raggedright, singlelinecheck=false}
\caption{Comparison between the results estimated using the
    zero $\mathcal{O}(p^4)$ LEC assumption (scenario I) and nonanalytic-dominance approximation (scenario II). The scalar and spin polarizabilities are in units of
    $10^{-4}\,\mathrm{fm}^{3}$ and $10^{-4}\,\mathrm{fm}^{4}$,
    respectively.}
\begin{ruledtabular}
\renewcommand{\arraystretch}{1.7}
\begin{tabular}{ccccccc}
    & $\Sigma_c^{++}$ & $\Sigma_c^{+}$ & $\Sigma_c^{0}$ & $\Xi_c^{\prime +}$ & $\Xi_c^{\prime 0}$ & $\Omega_c^{0}$ \\ \hline
    $\alpha_E^{\rm I}$ & 5.12 & 9.63 & 4.88 & 4.34 & 3.02 & 1.16 \\
    $\alpha_E^{\rm II}$ & 4.47 & 9.27 & 4.76 & 4.03 & 2.92 & 1.08 \\ \hline
    $\beta_M^{\rm I}$ & 7.74 & 0.44 & 0.28 & 0.48 & 0.09 & $-0.10$ \\
    $\beta_M^{\rm II}$ & 8.65 & 0.49 & $-0.18$ & 0.20 & $-0.77$ & $-1.35$ \\ \hline
    $\gamma_{E1E1}^{\rm I}$ & $-0.32$ & $-2.29$ & $-1.25$ & $-0.50$ & $-0.66$ & $-0.06$ \\
    $\gamma_{E1E1}^{\rm II}$ & $-1.64$ & $-2.52$ & $-1.20$ & $-0.81$ & $-0.63$ & $-0.06$ \\ \hline
    $\gamma_{M1M1}^{\rm I}$ & 0.72 & 0.02 & $-5.47$ & 0.85 & $-5.45$ & $-5.39$ \\
    $\gamma_{M1M1}^{\rm II}$ & 2.04 & 2.20 & $-6.07$ & 0.50 & $-6.72$ & $-7.47$ \\ \hline
    $\gamma_{E1M2}^{\rm I}$ & 0.26 & 0.46 & 0.21 & 0.13 & 0.11 & 0.01 \\
    $\gamma_{E1M2}^{\rm II}$ & 0.24 & 0.48 & 0.24 & 0.14 & 0.13 & 0.01 \\ \hline
    $\gamma_{M1E2}^{\rm I}$ & 0.66 & 0.47 & $-0.11$ & 0.15 & $-0.08$ & $-0.06$ \\
    $\gamma_{M1E2}^{\rm II}$ & 0.73 & 0.54 & $-0.09$ & 0.18 & $-0.07$ & $-0.06$ \\
\end{tabular}
\end{ruledtabular}
\label{tab10}
\end{table}

\section{\label{sec:5}SUMMARY}

Our work presents the first comprehensive calculation of the spin polarizabilities for spin-1/2 singly heavy baryons. Moreover, we have systematically extended the computation of both electromagnetic and spin polarizabilities to $\mathcal{O}(p^4)$, thereby achieving higher precision and examining the convergence properties of HBChPT in this context. 

Our results show that the electric polarizabilities of singly charmed baryons are smaller than those of nucleons. For the magnetic polarizability, the small mass splitting leads to a large contribution from $\mathcal{B}_{6^*}$, and we notice that the higher-order corrections to the magnetic polarizability are closely related to the transition magnetic moments. For the spin polarizabilities, higher-order corrections are nonnegligible similar to the nucleon case. We find that $\gamma_{M1M1}$ of singly heavy baryons is significantly larger compared to other spin polarizabilities. The properties exhibited by $\gamma_{E1E1}$ and $\gamma_{E1M2}$ in the calculation are similar to those of $\alpha_E$, while $\gamma_{M1M1}$ and $\gamma_{M1E2}$ are similar to $\beta_M$. 

This study is useful for the future experimental exploration of electromagnetic polarizabilities of singly heavy baryons. In this work we use the quark model, SU(4) symmetry, heavy quark symmetry, and so on to estimate the LECs, and the new experimental data and lattice QCD simulations would provide more accurate constraints on the LECs and verify our conclusions. In future, it is worthwhile to further investigate the generalized polarizabilities of singly heavy baryons via the virtual photon Compton scattering.

\begin{acknowledgments}
This work is supported by the National Natural Science Foundation of China under Grants No. 12335001, No. 12175091, No. 12247101, the “111 Center” under Grant No. B20063, and the innovation project for young science and technology talents of Lanzhou city under Grant No. 2023-QN-107. 
\end{acknowledgments}

\appendix

\section{Full Amplitude}
\label{appC}

\begin{table}
\captionsetup{justification=raggedright, singlelinecheck=false}
\caption{The values of coefficients $D_{\xi,\chi}$ for different mesons and single charmed baryons.}
\begin{ruledtabular}
\renewcommand{\arraystretch}{1.7}
\begin{tabular}{ccccccc}
 &$\Sigma_c^{++}$ & $\Sigma_c^{+}$ & $\Sigma_c^{0}$ & $\Xi_c^{'+}$ & $\Xi_c^{'0}$ & $\Omega_c^{0}$\\ \hline
 $D^{(c)}_\pi$ & $\frac{1}{2}$ & $1$ & $\frac{1}{2}$ & $\frac{1}{4}$ & $\frac{1}{4}$ & $0$ \\ 
 $D^{(c)}_K$ & $\frac{1}{2}$ & $\frac{1}{4}$ & $0$ & $1$ & $\frac{1}{4}$ & $\frac{1}{2}$ \\ 
 $D^{(d+e)}_\pi$ & $-2$ & $-4$ & $-2$ & $-1$ & $-1$ & $0$ \\ 
 $D^{(d+e)}_K$ & $-2$ & $-1$ & $0$ & $-4$ & $-1$ & $-2$ \\ 
 $D^{(f)}_\pi$ & $2$ & $4$ & $2$ & $1$ & $1$ & $0$ \\ 
 $D^{(f)}_K$ & $2$ & $1$ & $0$ & $4$ & $1$ & $2$ \\ 
 $D^{(g)}_\pi$ & $-\frac{1}{2}$ & $-1$ & $-\frac{1}{2}$ & $-\frac{1}{4}$ & $-\frac{1}{4}$ & $0$ \\ 
 $D^{(g)}_K$ & $-\frac{1}{2}$ & $-\frac{1}{4}$ & $0$ & $-1$ & $-\frac{1}{4}$ & $-\frac{1}{2}$ \\ \hline
 $D^{(c_3)}_{\pi}$ & $\frac{3}{2}$ & $-1$ & $-\frac{1}{2}$ & $\frac{1}{4}$ & $-\frac{1}{4}$ & $0$ \\ 
 $D^{(c_3)}_{K}$ & $\frac{3}{2}$ & $\frac{1}{4}$ & $0$ & $-1$ & $-\frac{1}{4}$ & $-\frac{1}{2}$ \\ 
 $D^{(d_3+e_3)}_{\pi}$ & $-6$ & $4$ & $2$ & $-1$ & $1$ & $0$ \\ 
 $D^{(d_3+e_3)}_{K}$ & $-6$ & $-1$ & $0$ & $4$ & $1$ & $2$ \\ 
\end{tabular}
\end{ruledtabular}
\label{tab7}
\end{table}

\begin{table}
\captionsetup{justification=raggedright, singlelinecheck=false}
\caption{The different values of coefficients $D_{\xi,\chi}$ for different mesons and single bottom baryons.}
\begin{ruledtabular}
\renewcommand{\arraystretch}{1.7}
\begin{tabular}{ccccccc}
 &$\Sigma_b^{+}$ & $\Sigma_b^{0}$ & $\Sigma_b^{-}$ & $\Xi_b^{'0}$ & $\Xi_b^{'-}$ & $\Omega_b^{-}$\\ \hline
 $D^{(c_3)}_{\pi}$ & $\frac{1}{2}$ & $-1$ & $\frac{1}{2}$ & $-\frac{1}{4}$ & $\frac{1}{4}$ & $0$ \\ 
 $D^{(c_3)}_{K}$ & $\frac{1}{2}$ & $-\frac{1}{4}$ & $0$ & $-1$ & $\frac{1}{4}$ & $\frac{1}{2}$ \\
 $D^{(d_3+e_3)}_{\pi}$ & $-2$ & $4$ & $-2$ & $1$ & $-1$ & $0$ \\ 
 $D^{(d_3+e_3)}_{K}$ & $-2$ & $1$ & $0$ & $4$ & $-1$ & $-2$ \\ 
\end{tabular}
\end{ruledtabular}
\label{tab8}
\end{table}

The full amplitudes for all $\mathcal{O}(p^3)$ and $\mathcal{O}(p^4)$ Feynman diagrams are listed below. For $\mathcal{O}(p^3)$ diagrams $a - g'$, since Ref.~\cite{Chen:2024xks} only provides the results of $A_{1,2}$, we provide here the complete results including the spin parts. We define $A = e^2 g_1^2 / 2$ and $B = e^2 g_3^2 / 2$. The auxiliary masses $\mathcal{M}_\chi$ and $\mathcal{M}'_\chi$ are
\begin{equation}
\begin{aligned}
\mathcal{M}_\chi =& M^2_\chi + 2xy\omega^2(1-\cos\theta), \\
\mathcal{M}'_\chi =& M^2_\chi + 2x(x-1)\omega^2(1-\cos\theta).
\end{aligned}
\label{eqc.1}
\end{equation}
We also need $D_{\xi,\chi}$ whose values can be found in Table~\ref{tab7} and \ref{tab8} for singly charmed and bottom baryons, respectively. For the linear combination of $c_i$  we define
\begin{equation*}
\begin{aligned}
    c_{\Sigma^{++}_c,\pi,1} =& 2c_0 + \frac{2(2m_d - m_u)}{3(m_d+m_u)}c_1, \\
    c_{\Sigma^{+}_c,\pi,1} =& 2c_0 + \frac{1}{3}c_1, \\
    c_{\Sigma^{0}_c,\pi,1} =& 2c_0 - \frac{2(m_d - 2m_u)}{3(m_d+m_u)}c_1, \\
    c_{\Xi^{+}_c,\pi,1} =& 2c_0 + \frac{m_d - 2m_u}{3(m_d + m_u)}c_1, \\
    c_{\Xi^{0}_c,\pi,1} =& 2c_0 - \frac{2m_d - m_u}{3(m_d + m_u)}c_1, \\
    c_{\Omega^{0}_c,\pi,1} =& 2c_0 - \frac{2}{3}c_1, \\
\end{aligned}
\end{equation*}
\begin{equation}
\begin{aligned}
    c_{\Sigma^{++}_c,K,1} =& 2c_0 + \frac{2(2m_s - m_u)}{3(m_s + m_u)}c_1, \\
    c_{\Sigma^{+}_c,K,1} =& 2c_0 + \frac{m_s - 2m_u}{3(m_s + m_u)}c_1, \\
    c_{\Sigma^{0}_c,K,1} =& 2c_0 - \frac{2}{3}c_1, \\
    c_{\Xi^{+}_c,K,1} =& 2c_0 + \frac{1}{3}c_1, \\
    c_{\Xi^{0}_c,K,1} =& 2c_0 - \frac{2m_s - m_u}{3(m_s + m_u)}c_1, \\
    c_{\Omega^{0}_c,K,1} =& 2c_0 - \frac{2(m_s - 2m_u)}{3(m_s + m_u)}c_1, \\
\end{aligned}
\label{eqc1.1}
\end{equation}
\begin{equation}
\begin{aligned}
    c_{\Sigma^{++}_c,\pi,2} =& c_{\Sigma^{0}_c,\pi,2} = c_{\Sigma^{++}_c,K,2} = c_{\Omega^{0}_c,K,2} \\
    =& \frac{1}{2}\left(c_2 - \frac{2g_2^2 + g_1^2}{4M_6}\right) + c_4, \\
    c_{\Xi^{+}_c,\pi,2} =& c_{\Xi^{0}_c,\pi,2} = c_{\Sigma^{+}_c,K,2} = c_{\Xi^{0}_c,K,2} \\
    =& \frac{1}{4}\left(c_2 - \frac{2g_2^2 + g_1^2}{4M_6}\right) + c_4, \\
    c_{\Sigma^{+}_c,\pi,2} =& c_{\Xi^{+}_c,K,2} = \frac{1}{2}\left(c_2 + c_3 - \frac{g_1^2}{2M_6}\right) + c_4, \\
    c_{\Omega^{0}_c,\pi,2} =& c_{\Sigma^{0}_c,K,2} = c_4, \\
\end{aligned}
\label{eqc1.2}
\end{equation}
For computational convenience in diagrams $f_2$ and $g_2$, we introduce a parameter $a$ and subsequently take the limit $a \to 0$. 

The meson-baryon loop integrals $J_i(\omega, M_\chi^2)$ and $\mathcal{J}_i(\omega,\delta,M_\chi^2)$ are defined in Refs.~\cite{Chen:2024xks, Peset:2014jxa}, the explicit forms of the corresponding meson-meson loop integrals $I_i(M_\chi^2)$ are given in Ref.~\cite{Scherer:2002tk}, and our function $\Delta (M_\chi^2)$ is identical to $\Delta_\chi$ in Ref.~\cite{Chen:2024xks}. For computational convenience, we define the following $\mathcal{G}$ function:
\begin{equation}
\begin{aligned}
&\mathcal{G}_i(\omega,\delta,M_\chi^2) = J_i(\omega-\delta,M_\chi^2) - J_i(-\omega-\delta,M_\chi^2). \\
\end{aligned}
\label{eqb.6}
\end{equation}
The $\mathcal{G}^{\prime}_i$ and $\mathcal{G}^{\prime \prime}_i$ denote the first and second partial derivatives of $\mathcal{G}_i$ with respect to $M^2_\chi$, respectively.

\subsection{$\mathcal{O}(p^3)$ diagrams with spin-1/2 contribution}

For diagram $a$:
\begin{equation}
\begin{aligned}
    A_{1,\xi } = -\frac{e^2 Q^2_\xi}{M_6}.
\end{aligned}
\label{eqc.2}
\end{equation}

For diagram $b$, the s-channel and u-channel contributions cancel, resulting in 0.

For diagram $c$:
\begin{equation}
\begin{aligned}
A_{1,\xi } =& A \sum_\chi \frac{D^{(c)}_{\xi,\chi}}{F_\chi^2}\mathcal{J}_0(\omega ,0,M_\chi^2), 
\end{aligned}
\label{eqc.3}
\end{equation}
\begin{equation}
\begin{aligned}
A_{3,\xi } =& A \sum_\chi \frac{D^{(c)}_{\xi,\chi}}{F_\chi^2}\mathcal{G}_0(\omega ,0,M_\chi^2).
\end{aligned}
\label{eqc.4}
\end{equation}

For diagram $(d+e)$:
\begin{equation}
\begin{aligned}
A_{1,\xi } =& A \sum_\chi \frac{D^{(d+e)}_{\xi,\chi}}{F_\chi^2}\int_0^1 dx \mathcal{J}_2^\prime(x\omega,0,M_\chi^2), 
\end{aligned}
\label{eqc.5}
\end{equation}
\begin{equation}
\begin{aligned}
A_{3,\xi } =& A \sum_\chi \frac{D^{(d+e)}_{\xi,\chi}}{F_\chi^2}\int_0^1 dx \mathcal{G}_2^\prime(x\omega,0,M_\chi^2), 
\end{aligned}
\label{eqc.7}
\end{equation}

For diagram $f$:
\begin{equation}
\begin{aligned}
A_{1,\xi } =& A \sum_\chi \frac{D^{(f)}_{\xi,\chi}}{F_\chi^2}\int_0^1 dx \int_0^{1-x} dy \left\{\omega^2 \left[-x^2 - y^2 \right.\right. \\
& \left. - 2xy\cos\theta  + (x+y)(1+\cos\theta)-\cos\theta\right] \\
& \mathcal{J}_2^{\prime \prime}\left((1-x-y)\omega ,0,\mathcal{M}_\chi\right) \\
& \left.  + (d+1) \mathcal{J}_6^{\prime \prime}\left((1-x-y)\omega ,0,\mathcal{M}_\chi\right)\right\}, 
\end{aligned}
\label{eqc.9}
\end{equation}
\begin{equation}
\begin{aligned}
A_{2,\xi } =& A \sum_\chi \frac{D^{(f)}_{\xi,\chi}}{F_\chi^2}\int_0^1 dx \int_0^{1-x} dy \omega^2 \left[-(3+d)xy\right. \\
&\left.+ x + y \right] \mathcal{J}_2^{\prime \prime}\left((1-x-y)\omega ,0,\mathcal{M}_\chi\right), 
\end{aligned}
\label{eqc.10}
\end{equation}
\begin{equation}
\begin{aligned}
A_{4,\xi } =& A \sum_\chi \frac{D^{(f)}_{\xi,\chi}}{F_\chi^2}\int_0^1 dx \int_0^{1-x} dy \omega^2 (x+y-1) \\
&\mathcal{G}_2^{\prime \prime}\left((1-x-y)\omega ,0,\mathcal{M}_\chi\right), 
\end{aligned}
\label{eqc.11}
\end{equation}
\begin{equation}
\begin{aligned}
A_{5,\xi } =& A \sum_\chi \frac{D^{(f)}_{\xi,\chi}}{F_\chi^2}\int_0^1 dx \int_0^{1-x} dy \omega^2 y \\
&\mathcal{G}_2^{\prime \prime}\left((1-x-y)\omega ,0,\mathcal{M}_\chi\right), 
\end{aligned}
\label{eqc.12}
\end{equation}
\begin{equation}
\begin{aligned}
A_{6,\xi } =& A \sum_\chi \frac{D^{(f)}_{\xi,\chi}}{F_\chi^2}\int_0^1 dx \int_0^{1-x} dy \omega^2 \left(-y\right) \\
&\mathcal{G}_2^{\prime \prime}\left((1-x-y)\omega ,0,\mathcal{M}_\chi\right).
\end{aligned}
\label{eqc.13}
\end{equation}

For diagram $g$:
\begin{equation}
\begin{aligned}
A_{1,\xi} =& A \sum_\chi \frac{D^{(g)}_{\xi,\chi} }{F^2_\chi}\int_0^1 dx \left[(d-1)\mathcal{J}^\prime_2(0,0,\mathcal{M}'_\chi)\right. \\
& \left. + 2\omega^2 (1 - \cos\theta )x(1-x) \mathcal{J}^\prime_0(0,0,\mathcal{M}'_\chi)\right].
\end{aligned}
\label{eqc.14}
\end{equation}

\subsection{$\mathcal{O}(p^4)$ diagrams with spin-1/2 contribution}

For diagram $c_2$:
\begin{equation}
\begin{aligned}
A_{1,\xi} =& A \sum_\chi \frac{D^{(c)}_{\xi,\chi}}{2M_6 F^2_\chi}\left[-2\Delta (M_\chi^2) + \frac{\partial}{\partial \omega}\mathcal{G}_3(\omega ,0, M_\chi ^2)\right. \\
&\left. + d\frac{\partial}{\partial \omega}\mathcal{G}_2(\omega ,0, M_\chi ^2) + 2\omega \frac{\partial}{\partial \omega}\mathcal{J}_1(\omega ,0, M_\chi ^2)\right], 
\end{aligned}
\label{eqc.15}
\end{equation}
\begin{equation}
\begin{aligned}
A_{3,\xi} =& A \sum_\chi \frac{D^{(c)}_{\xi,\chi}}{2M_6 F^2_\chi}\left[\frac{\partial}{\partial \omega}\mathcal{J}_3(\omega ,0, M_\chi ^2) \right. \\
&\left. + d\frac{\partial}{\partial \omega}\mathcal{J}_2(\omega ,0, M_\chi ^2) + 2\omega \frac{\partial}{\partial \omega}\mathcal{G}_1(\omega ,0, M_\chi ^2)\right].
\end{aligned}
\label{eqc.16}
\end{equation}

For diagram $c_3$:
\begin{equation}
\begin{aligned}
A_{1,\xi} = A \sum_\chi \frac{D^{(c_3)}_{\xi,\chi} }{F_\chi^2 M_6}\mathcal{J}_1(\omega,0,M_\chi^2),
\end{aligned}
\label{eqc.17}
\end{equation}
\begin{equation}
\begin{aligned}
A_{3,\xi} = A \sum_\chi \frac{D^{(c_3)}_{\xi,\chi} }{F_\chi^2 M_6}\mathcal{G}_1(\omega,0,M_\chi^2).
\end{aligned}
\label{eqc.18}
\end{equation}

For diagram $c_4$:
\begin{equation}
\begin{aligned}
A_{1,\xi} = A \sum_\chi \frac{D^{(c)}_{\xi,\chi} }{F_\chi^2 M_6}\frac{2Q_\xi}{\omega}\mathcal{G}_2(\omega,0,M_\chi^2),
\end{aligned}
\end{equation}
\begin{equation}
\begin{aligned}
A_{3,\xi} = A \sum_\chi \frac{D^{(c)}_{\xi,\chi} }{F_\chi^2 M_6}\frac{2Q_\xi}{\omega}\left[\mathcal{J}_2(\omega,0,M_\chi^2) - \mathcal{J}_2(0,0,M_\chi^2)\right].
\end{aligned}
\end{equation}

For diagram $(d_2+e_2)$:
\begin{equation}
\begin{aligned}
A_{1,\xi} =& A\sum_\chi \frac{D^{(d+e)}_{\xi,\chi}}{2M_6 F^2_\chi}\left\{2I_2(M_\chi^2) + \int_0^1 dx \frac{1}{1-x}\right. \\
&\left[\frac{\partial}{\partial \omega}\mathcal{G}_7^\prime((1-x)\omega,0,M_\chi^2)\right. \\
&+(d+2)\frac{\partial}{\partial \omega}\mathcal{G}_6^\prime((1-x)\omega,0,M_\chi^2) \\
&+2\omega(1-x)\frac{\partial}{\partial \omega}\mathcal{J}_4^\prime((1-x)\omega,0,M_\chi^2) \\
&\left.\left.-\omega^2(1-\cos\theta)x\frac{\partial}{\partial \omega}\mathcal{G}_2^\prime((1-x)\omega,0,M_\chi^2)\right]\right\},
\end{aligned}
\label{eqc.19}
\end{equation}
\begin{equation}
\begin{aligned}
A_{2,\xi} =& - A\sum_\chi \frac{D^{(d+e)}_{\xi,\chi}}{2M_6 F^2_\chi}\omega^2\int_0^1 dx \\
&\frac{\partial}{\partial \omega}\mathcal{G}_2^\prime((1-x)\omega,0,M_\chi^2),
\end{aligned}
\label{eqc.20}
\end{equation}
\begin{equation}
\begin{aligned}
A_{3,\xi} =& A\sum_\chi \frac{D^{(d+e)}_{\xi,\chi}}{2M_6 F^2_\chi}\left\{\int_0^1 dx \frac{1}{1-x}\right. \\
& \left[\frac{\partial}{\partial \omega}\mathcal{J}_7^\prime((1-x)\omega,0,M_\chi^2)\right. \\
&+(d+2)\frac{\partial}{\partial \omega}\mathcal{J}_6^\prime((1-x)\omega,0,M_\chi^2) \\
&+2\omega(1-x)\frac{\partial}{\partial \omega}\mathcal{G}_4^\prime((1-x)\omega,0,M_\chi^2) \\
&\left.\left.-\omega^2(1-\cos\theta)x\frac{\partial}{\partial \omega}\mathcal{J}_2^\prime((1-x)\omega,0,M_\chi^2)\right]\right\},
\end{aligned}
\label{eqc.21}
\end{equation}
\begin{equation}
\begin{aligned}
A_{5,\xi} =& - A\sum_\chi \frac{D^{(d+e)}_{\xi,\chi}}{2M_6 F^2_\chi} \omega^2\int_0^1 dx \\
&\frac{\partial}{\partial \omega}J_2^\prime(-(1-x)\omega,M_\chi^2),
\end{aligned}
\label{eqc.22}
\end{equation}

For diagram $(d_3+e_3)$:
\begin{equation}
\begin{aligned}
A_{1,\xi} =& A \sum_\chi \frac{D^{(d_3+e_3)}_{\xi,\chi} }{F_\chi^2 M_6} \left[I_2(M_\chi^2) - \int_0^1 dx \right.\\
&\left. \frac{\omega}{2} \mathcal{G}^\prime_2(x\omega,0,M^2_\chi)\right],
\end{aligned}
\label{eqc.24}
\end{equation}
\begin{equation}
\begin{aligned}
A_{3,\xi} =& A \sum_\chi \frac{D^{(d_3+e_3)}_{\xi,\chi} }{F_\chi^2 M_6} \int_0^1 dx \left( -\frac{\omega}{2}\right) \mathcal{J}^\prime_2(x\omega,0,M^2_\chi),
\end{aligned}
\label{eqc.26}
\end{equation}

For diagram $(d_4+e_4)$:
\begin{equation}
\begin{aligned}
A_{1,\xi} = - A \sum_\chi \frac{D^{(d+e)}_{\xi,\chi}}{F_\chi^2 M_6} I_2(M^2_\chi),
\end{aligned}
\label{eqc.28}
\end{equation}

For diagram $(d_5+e_5)$:
\begin{equation}
\begin{aligned}
A_{1,\xi} =& A\sum_\chi \frac{D^{(d+e)}_{\xi,\chi} C_{2,\xi}}{2M_N F^2_\chi}\int_0^1 dx \left(-\omega \cos\theta \right) \\
& \mathcal{G}^\prime_2\left((1-x)\omega,0,M^2_\chi\right),
\end{aligned}
\label{eqc.30}
\end{equation}
\begin{equation}
\begin{aligned}
A_{2,\xi} =& A\sum_\chi \frac{D^{(d+e)}_{\xi,\chi} C_{2,\xi}}{2M_N F^2_\chi}\int_0^1 dx \omega \mathcal{G}^\prime_2\left((1-x)\omega,0,M^2_\chi\right),
\end{aligned}
\label{eqc.31}
\end{equation}
\begin{equation}
\begin{aligned}
A_{3,\xi} =& A\sum_\chi \frac{D^{(d+e)}_{\xi,\chi} C_{2,\xi}}{2M_N F^2_\chi}\int_0^1 dx \left(-\omega \cos\theta \right) \\
&\mathcal{J}^\prime_2\left((1-x)\omega,0,M^2_\chi\right),
\end{aligned}
\label{eqc.32}
\end{equation}
\begin{equation}
\begin{aligned}
A_{4,\xi} =& A\sum_\chi \frac{D^{(d+e)}_{\xi,\chi} C_{2,\xi}}{2M_N F^2_\chi}\int_0^1 dx \left(-\omega \right) \\
&\mathcal{J}^\prime_2\left((1-x)\omega,0,M^2_\chi\right),
\end{aligned}
\label{eqc.33}
\end{equation}
\begin{equation}
\begin{aligned}
A_{5,\xi} =& A\sum_\chi \frac{D^{(d+e)}_{\xi,\chi} C_{2,\xi}}{2M_N F^2_\chi}\int_0^1 dx \omega \\
& \mathcal{J}^\prime_2\left((1-x)\omega,0,M^2_\chi\right).
\end{aligned}
\label{eqc.34}
\end{equation}
\begin{equation}
\begin{aligned}
A_{6,\xi} =& A\sum_\chi \frac{D^{(d+e)}_{\xi,\chi} Q_\xi}{2M_6 F^2_\chi}\int_0^1 dx (-\omega) \\
& J^\prime_2\left(-(1-x)\omega,M^2_\chi\right).
\end{aligned}
\end{equation}

For diagram $f_2$:
\begin{equation}
\begin{aligned}
A_{1,\xi} =& A \sum_\chi \frac{D^{(f)}_{\xi ,\chi}}{2F_\chi^2 M_6}\int_0^1 dx \left\{\int_0^{1-x} dy \left\{(d+1) \right. \right.\\
&\left[2I_6^\prime(\mathcal{M}_\chi) - M^2_\chi \frac{\partial}{\partial a}\mathcal{G}_6^{\prime \prime}((1-x-y)\omega - a,0,\mathcal{M}_\chi)\right] \\
&  - \omega^2 \left[x^2+y^2+2xy\cos\theta - (x+y)(1+\cos\theta) \right.\\
&\left.+\cos\theta\right] \left[2I_2^\prime(\mathcal{M}_\chi) - M^2_\chi \frac{\partial}{\partial a}\right.\\
&\left.\left. \mathcal{G}_2^{\prime \prime}((1-x-y)\omega - a,0,\mathcal{M}_\chi)\right]\right\} \\
& + \left\{\frac{d+1}{1-x}\frac{\partial}{\partial \omega}\mathcal{G}_6^\prime((1-x)\omega,0,M^2_\chi) - \frac{1}{1-x}\omega^2\right. \\
& \left[x^2 - x(1+\cos\theta)+\cos\theta\right] \\
& \left.\left.\frac{\partial}{\partial \omega}\mathcal{G}_2^\prime((1-x)\omega,0,M^2_\chi)\right\}\right\},
\end{aligned}
\label{eqc.35}
\end{equation}
\begin{equation}
\begin{aligned}
A_{2,\xi} =& A \sum_\chi \frac{D^{(f)}_{\xi ,\chi}}{2F_\chi^2 M_6} \int_0^1 dx \omega^2\left\{ \int_0^{1-x} dy \right. \\
&\left[x+y-(3+d)xy\right] \\
& \left[2I_2^\prime(\mathcal{M}_\chi) - M^2_\chi \frac{\partial}{\partial a}\mathcal{G}_2^{\prime \prime}((1-x-y)\omega - a,0,\mathcal{M}_\chi)\right] \\
& + \frac{x}{1-x}\left. \frac{\partial}{\partial \omega}\mathcal{G}^\prime_2((1-x)\omega,0,M^2_\chi) \right\},
\end{aligned}
\label{eqc.36}
\end{equation}
\begin{equation}
\begin{aligned}
A_{4,\xi} =& A \sum_\chi \frac{D^{(f)}_{\xi ,\chi}}{2F_\chi^2 M_6}\int_0^1 dx \omega^2\left[ \int_0^{1-x} dy (-x-y+1)\right. \\
& M^2_\chi \frac{\partial}{\partial a}\mathcal{J}_2^{\prime \prime}((1-x-y)\omega - a,0,\mathcal{M}_\chi) \\
&\left. - \frac{\partial}{\partial \omega}\mathcal{J}^\prime_2((1-x)\omega,0,M^2_\chi)\right],
\end{aligned}
\label{eqc.37}
\end{equation}
\begin{equation}
\begin{aligned}
A_{5,\xi} = & A \sum_\chi \frac{D^{(f)}_{\xi ,\chi}}{2F_\chi^2 M_6} \int_0^1 dx \omega^2\left[\int_0^{1-x} dy (-x)\right. \\
& M^2_\chi \frac{\partial}{\partial a}\mathcal{J}_2^{\prime \prime}((1-x-y)\omega - a,0,\mathcal{M}_\chi) \\
&\left. + \frac{x}{1-x} \frac{\partial}{\partial \omega} J^\prime_2((1-x)\omega,M^2_\chi)\right],
\end{aligned}
\label{eqc.38}
\end{equation}
\begin{equation}
\begin{aligned}
A_{6,\xi} =& A \sum_\chi \frac{D^{(f)}_{\xi ,\chi}}{2F_\chi^2 M_6} \int_0^1 dx \omega^2\left[\int_0^{1-x} dy \left(x\right)\right. \\
& M^2_\chi \frac{\partial}{\partial a}\mathcal{J}_2^{\prime \prime}((1-x-y)\omega - a,0,\mathcal{M}_\chi) \\
&\left. - \frac{x}{1-x} \frac{\partial}{\partial \omega} J^\prime_2((1-x)\omega,M^2_\chi) \right].
\end{aligned}
\label{eqc.39}
\end{equation}

For diagram $f_3$:
\begin{equation}
\begin{aligned}
A_{1,\xi} =& A \sum_\chi \frac{D^{(f)}_{\xi ,\chi}}{2F_\chi^2 M_6}\int_0^1 dx \int_0^{1-x} dy \left\{-4(d+1) I_6^\prime(\mathcal{M}_\chi)\right. \\
& + 4\omega^2 \left[x^2+y^2+2xy\cos\theta-(x + y)(1 + \cos\theta)\right. \\
&\left.\left.+1\right] I_2^\prime(\mathcal{M}_\chi) \right\},
\end{aligned}
\label{eqc.40}
\end{equation}
\begin{equation}
\begin{aligned}
A_{2,\xi} =& A \sum_\chi \frac{D^{(f)}_{\xi ,\chi}}{2F_\chi^2 M_6}\int_0^1 dx \int_0^{1-x} dy \omega^2 \\
& 4\left[-x-y+(d+3)xy\right] I_2^\prime(\mathcal{M}_\chi).
\end{aligned}
\label{eqc.41}
\end{equation}

For diagram $f_4$:
\begin{equation}
\begin{aligned}
A_{1,\xi} =& A\sum_\chi \frac{D^{(f)}_{\xi,\chi} \tilde{C}_{2,\xi,\chi}}{2M_N F^2_\chi}\int_0^1 dx \frac{\omega \cos\theta }{2} \\
& \left[\mathcal{G}^\prime_2\left((1-x)\omega,0,M^2_\chi\right)-\mathcal{G}^\prime_2\left((-x)\omega,0,M^2_\chi\right)\right] \\
& + A\sum_\chi \frac{D^{(f)}_{\xi,\chi} Q_\xi}{2M_6 F^2_\chi}\int_0^1 dx \left\{\left(-\frac{d+1}{\omega}\right)\right. \\
& \left[\mathcal{G}^\prime_6\left((1-x)\omega,0,M^2_\chi\right)-\mathcal{G}^\prime_6\left((-x)\omega,0,M^2_\chi\right)\right] \\
& + \omega x(x-1) \left[\mathcal{G}^\prime_2\left((1-x)\omega,0,M^2_\chi\right)\right. \\
&\left. \left.-\mathcal{G}^\prime_2\left((-x)\omega,0,M^2_\chi\right)\right]\right\},
\end{aligned}
\label{eqc.42}
\end{equation}
\begin{equation}
\begin{aligned}
A_{2,\xi} =& A\sum_\chi \frac{D^{(f)}_{\xi,\chi} \tilde{C}_{2,\xi,\chi}}{2M_N F^2_\chi}\int_0^1 dx \left(-\frac{\omega}{2}\right) \\
& \left[\mathcal{G}^\prime_2\left((1-x)\omega,0,M^2_\chi\right)-\mathcal{G}^\prime_2\left((-x)\omega,0,M^2_\chi\right)\right],
\end{aligned}
\label{eqc.43}
\end{equation}
\begin{equation}
\begin{aligned}
A_{3,\xi} =& A\sum_\chi \frac{D^{(f)}_{\xi,\chi} \tilde{C}_{2,\xi,\chi}}{2M_N F^2_\chi}\int_0^1 dx \left(-\frac{\omega \cos\theta }{2}\right)(1-2x) \\
& \left[\mathcal{J}^\prime_2\left((1-x)\omega,0,M^2_\chi\right)-\mathcal{J}^\prime_2\left((-x)\omega,0,M^2_\chi\right)\right],
\end{aligned}
\label{eqc.44}
\end{equation}
\begin{equation}
\begin{aligned}
A_{4,\xi} =& A\sum_\chi \frac{D^{(f)}_{\xi,\chi} \tilde{C}_{2,\xi,\chi}}{2M_N F^2_\chi}\int_0^1 dx \left(-\frac{\omega}{2}\right)(2x-1) \\
& \left[\mathcal{J}^\prime_2\left((1-x)\omega,0,M^2_\chi\right)-\mathcal{J}^\prime_2\left((-x)\omega,0,M^2_\chi\right)\right],
\end{aligned}
\label{eqc.45}
\end{equation}
\begin{equation}
\begin{aligned}
A_{5,\xi} =& A\sum_\chi \frac{D^{(f)}_{\xi,\chi} \tilde{C}_{2,\xi,\chi}}{2M_N F^2_\chi}\int_0^1 dx \frac{\omega}{2}(2x-1) \\
&\left[\mathcal{G}^\prime_2\left((1-x)\omega,0,M^2_\chi\right)-\mathcal{G}^\prime_2\left((-x)\omega,0,M^2_\chi\right)\right].
\end{aligned}
\label{eqc.46}
\end{equation}

\begin{equation}
\begin{aligned}
A_{6,\xi} =& A\sum_\chi \frac{D^{(f)}_{\xi,\chi} Q_\xi}{2M_6 F^2_\chi}\int_0^1 dx \omega(x-1) \\
&\left[J^\prime_2\left(-(1-x)\omega,M^2_\chi\right)-J^\prime_2\left(-(-x)\omega,M^2_\chi\right)\right].
\end{aligned}
\end{equation}

For diagram $g_2$:
\begin{equation}
\begin{aligned}
A_{1,\xi} =& A \sum_\chi \frac{D^{(g)}_{\xi,\chi}}{2M_6 F^2_\chi} \int_0^1 dx \left[2(d-1)I_2(\mathcal{M}^\prime_\chi)\right. \\
& + 4x(1-x)\omega^2 (1-\cos\theta)I_0(\mathcal{M}^\prime_\chi) \\
& - 2(d-1)\frac{\partial}{\partial a}J^\prime_7(-a,\mathcal{M}^\prime_\chi) \\
& - 4(d+5) \frac{\partial}{\partial a}J^\prime_6(-a,\mathcal{M}^\prime_\chi) \\
& + 4x(x-1)\omega^2(1-\cos\theta)\frac{\partial}{\partial a}J^\prime_3(-a,\mathcal{M}^\prime_\chi) \\
& \left. + x\left(44x-24\right)\omega^2(1-\cos\theta)\frac{\partial}{\partial a}J^\prime_2(-a,\mathcal{M}^\prime_\chi)\right].
\end{aligned}
\label{eqc.47}
\end{equation}

For diagram $g_3$:
\begin{equation}
\begin{aligned}
A_{1,\xi} =& A \sum_\chi \frac{D^{(g)}_{\xi,\chi}}{M_6 F^2_\chi}\int_0^1 dx \left[2(d-1) I_2(\mathcal{M}^\prime_\chi)\right. \\
&\left. + 4 \omega^2 x^2 (\cos\theta - 1)I_0(\mathcal{M}^\prime_\chi)\right].
\end{aligned}
\label{eqc.48}
\end{equation}

For diagrams $h_2$ and $h_3$, the results are 0.

For diagram $h_4$:
\begin{equation}
\begin{aligned}
A_{1,\xi} =& \sum_\chi \frac{e^2 }{F^2_\chi}\int_0^1 dx \int_0^{1-x} dy \\
&\left\{8 c_{\xi,\chi,2}I'_6(\mathcal{M}_\chi) + 8 \omega^2 \right.\\
&\left. c_{\xi,\chi,2} \left[x^2+y^2-2(x+y)-2xy+1\right] I'_2(\mathcal{M}_\chi)\right\} \\
& - \sum_\chi \frac{e^2 M^2_\chi c_{\xi,\chi,1}}{F_\chi^2} \int_0^1 dx \int_0^{1-x} dy 8I'_2(\mathcal{M}_\chi),
\end{aligned}
\label{eqc.126}
\end{equation}

\begin{equation}
\begin{aligned}
A_{2,\xi} =& - \sum_\chi \frac{e^2 c_{\xi,\chi,2}}{F^2_\chi}\omega^2 \int_0^1 dx \int_0^{1-x} dy\\
& \left[8(x^2 + y^2) - 8(x+y) +16xy + 2\right] I'_2(\mathcal{M}_\chi) \\
& + \sum_\chi \frac{e^2 M^2_\chi c_{\xi,\chi,1}}{F_\chi^2} \omega^2 \int_0^1 dx \int_0^{1-x} dy \\
&\left[8(x^2+y^2)-8(x+y)+16xy+2\right] I'_0(\mathcal{M}_\chi).
\end{aligned}
\label{eqc.127}
\end{equation}

For diagram $h_5$:
\begin{equation}
\begin{aligned}
A_{1,\xi} =& - \sum_\chi \frac{e^2 c_{\xi,\chi,2}}{F_\chi^2} \int_0^1 dx 2 I_2(\mathcal{M}'_\chi)  \\
& + \sum_\chi \frac{e^2 M^2_\chi c_{\xi,\chi,1}}{F^2_\chi} \int_0^1 dx 2I_0(\mathcal{M}'_\chi).
\end{aligned}
\end{equation}

\subsection{$\mathcal{O}(p^3)$ diagrams with spin-3/2 contribution}

For diagram $b'$:
\begin{equation}
\begin{aligned}
A_{1,\xi } = \frac{e^2C^2_{1,\xi}}{12M_N^2}\omega^2 \cos\theta \frac{\delta}{\delta^2-\omega^2},
\end{aligned}
\label{eqc.49}
\end{equation}
\begin{equation}
\begin{aligned}
A_{2,\xi } = -\frac{e^2C^2_{1,\xi}}{12M_N^2}\omega^2 \frac{\delta}{\delta^2-\omega^2},
\end{aligned}
\label{eqc.50}
\end{equation}
\begin{equation}
\begin{aligned}
A_{3,\xi } = \frac{e^2C^2_{1,\xi}}{24M_N^2}\omega^2 \cos\theta\frac{\omega}{\delta^2-\omega^2},
\end{aligned}
\label{eqc.51}
\end{equation}
\begin{equation}
\begin{aligned}
A_{4,\xi } = \frac{e^2C^2_{1,\xi}}{24M_N^2}\omega^2 \frac{\omega}{\delta^2-\omega^2},
\end{aligned}
\label{eqc.52}
\end{equation}
\begin{equation}
\begin{aligned}
A_{5,\xi } = -\frac{e^2C^2_{1,\xi}}{24M_N^2}\omega^2 \frac{\omega}{\delta^2-\omega^2}.
\end{aligned}
\label{eqc.53}
\end{equation}

For diagram $c'$:
\begin{equation}
\begin{aligned}
A_{1,\xi } = \frac{d-2}{d-1}B \sum_\chi \frac{D^{(c)}_{\xi,\chi}}{F_\chi^2}\mathcal{J}_0(\omega ,\delta,M_\chi^2),
\end{aligned}
\label{eqc.54}
\end{equation}
\begin{equation}
\begin{aligned}
A_{3,\xi } = - \frac{1}{d-1}B \sum_\chi \frac{D^{(c)}_{\xi,\chi}}{F_\chi^2}\mathcal{G}_0(\omega ,\delta,M_\chi^2).
\end{aligned}
\label{eqc.55}
\end{equation}

For diagram $(d'+e')$:
\begin{equation}
\begin{aligned}
A_{1,\xi } = \frac{d-2}{d-1}B \sum_\chi \frac{D^{(d+e)}_{\xi,\chi}}{F_\chi^2}\int_0^1 dx \mathcal{J}_2^\prime(x\omega,\delta,M_\chi^2),
\end{aligned}
\label{eqc.56}
\end{equation}
\begin{equation}
\begin{aligned}
A_{3,\xi } = -\frac{1}{d-1}B \sum_\chi \frac{D^{(d+e)}_{\xi,\chi}}{F_\chi ^2}\int_0^1 dx \mathcal{G}_2^\prime(x\omega,\delta,M_\chi^2),
\end{aligned}
\label{eqc.58}
\end{equation}

For diagram $f'$:
\begin{equation}
\begin{aligned}
A_{1,\xi } =& \frac{d-2}{d-1}B \sum_\chi \frac{D^{(f)}_{\xi,\chi}}{F_\chi^2}\int_0^1 dx \int_0^{1-x} dy \left\{\omega^2\right. \\
&\left[-x^2-2xy\cos\theta-y^2 +(x+y)(1+\cos\theta)-\cos\theta\right] \\
&\mathcal{J}_2^{\prime \prime}\left((1-x-y)\omega ,\delta,\mathcal{M}_\chi\right)\\
&\left. + (d+1) \mathcal{J}_6^{\prime \prime}\left((1-x-y)\omega ,\delta,\mathcal{M}_\chi\right)\right\},
\end{aligned}
\label{eqc.60}
\end{equation}
\begin{equation}
\begin{aligned}
A_{2,\xi } =& \frac{d-2}{d-1}B\sum_\chi \frac{D^{(f)}_{\xi,\chi}}{F_\chi^2}\int_0^1 dx \int_0^{1-x} dy \omega^2 \\
&\left[-(d+3)xy + x + y \right]\mathcal{J}_2^{\prime \prime}\left((1-x-y)\omega ,\delta,\mathcal{M}_\chi\right),
\end{aligned}
\label{eqc.61}
\end{equation}
\begin{equation}
\begin{aligned}
A_{4,\xi } =& -\frac{1}{d-1}B \sum_\chi \frac{D^{(f)}_{\xi,\chi}}{F_\chi^2}\int_0^1 dx \int_0^{1-x} dy \omega^2(x+y-1) \\
&\mathcal{G}_2^{\prime \prime}\left((1-x-y)\omega ,\delta,\mathcal{M}_\chi\right),
\end{aligned}
\label{eqc.62}
\end{equation}
\begin{equation}
\begin{aligned}
A_{5,\xi } =& -\frac{1}{d-1}B \sum_\chi \frac{D^{(f)}_{\xi,\chi}}{F_\chi^2}\int_0^1 dx \int_0^{1-x} dy \omega^2 y \\
&\mathcal{G}_2^{\prime \prime}\left((1-x-y)\omega ,\delta,\mathcal{M}_\chi\right),
\end{aligned}
\label{eqc.63}
\end{equation}
\begin{equation}
\begin{aligned}
A_{6,\xi } =& -\frac{1}{d-1}B \sum_\chi \frac{D^{(f)}_{\xi,\chi}}{F_\chi^2}\int_0^1 dx \int_0^{1-x} dy \omega^2(-y) \\
&\mathcal{G}_2^{\prime \prime}\left((1-x-y)\omega ,\delta,\mathcal{M}_\chi\right).
\end{aligned}
\label{eqc.64}
\end{equation}

For diagram $g'$:
\begin{equation}
\begin{aligned}
A_{1,\xi } =& \frac{d-2}{d-1}B \sum_\chi \frac{D^{(g)}_{\xi,\chi} }{F^2_\chi}\int_0^1 dx \left[(d-1)\mathcal{J}^\prime_2(0,\delta,\mathcal{M}^\prime_\chi)\right. \\
&\left. + 2\omega^2 (1 - \cos\theta )x(1-x) \mathcal{J}^\prime_0(0,\delta,\mathcal{M}^\prime_\chi)\right].
\end{aligned}
\label{eqc.65}
\end{equation}

\subsection{$\mathcal{O}(p^4)$ diagrams with spin-3/2 contribution}

For diagram $c'_2$:
\begin{equation}
\begin{aligned}
A_{1,\xi} =& \frac{d-2}{d-1}B \sum_\chi \frac{D^{(c)}_{\xi,\chi}}{2M_{6^*} F_\chi^2}\left[-2\Delta (M_\chi^2) + \frac{\partial}{\partial \omega}\mathcal{G}_3(\omega , \delta,M^2_\chi)\right. \\
& + d\frac{\partial}{\partial \omega}\mathcal{G}_2(\omega , \delta,M^2_\chi) + 2\omega\frac{\partial}{\partial \omega}\mathcal{J}_1(\omega , \delta,M^2_\chi) \\
& - 2\delta\frac{\partial}{\partial \omega}\mathcal{G}_1(\omega , \delta,M^2_\chi) + \delta^2 \frac{\partial}{\partial \omega}\mathcal{G}_0(\omega , \delta,M^2_\chi) \\
&\left. - 2\omega \delta\frac{\partial}{\partial \omega}\mathcal{J}_0(\omega , \delta,M^2_\chi)\right],
\end{aligned}
\label{eqc.66}
\end{equation}
\begin{equation}
\begin{aligned}
A_{3,\xi} =& - \frac{1}{d-1}B \sum_\chi \frac{D^{(c)}_{\xi,\chi}}{2M_{6^*}  F_\chi^2}\left[\frac{\partial}{\partial \omega}\mathcal{J}_3(\omega , \delta,M^2_\chi)\right. \\
& + d\frac{\partial}{\partial \omega}\mathcal{J}_2(\omega , \delta,M^2_\chi) + 2\omega\frac{\partial}{\partial \omega}\mathcal{G}_1(\omega , \delta,M^2_\chi) \\
& - 2\delta\frac{\partial}{\partial \omega}\mathcal{J}_1(\omega , \delta,M^2_\chi) + \delta^2 \frac{\partial}{\partial \omega}\mathcal{J}_0(\omega , \delta,M^2_\chi) \\
&\left. - 2\omega \delta\frac{\partial}{\partial \omega}\mathcal{G}_0(\omega , \delta,M^2_\chi)\right].
\end{aligned}
\label{eqc.67}
\end{equation}

For diagram $c'_3$:
\begin{equation}
\begin{aligned}
A_{1,\xi } = \frac{d-2}{d-1}B \sum_\chi \frac{D^{(c_3)}_{\xi,\chi} }{F_\chi^2 M_6}\mathcal{J}_1(\omega,\delta,M_\chi^2),
\end{aligned}
\label{eqc.68}
\end{equation}
\begin{equation}
\begin{aligned}
A_{3,\xi } = -\frac{1}{d-1}B \sum_\chi \frac{D^{(c_3)}_{\xi,\chi} }{F_\chi^2 M_6}\mathcal{G}_1(\omega,\delta,M_\chi^2).
\end{aligned}
\label{eqc.69}
\end{equation}

For diagrams $c'_4$ and $c''_4$, their amplitudes are $0$.

For diagram $c'''_4$:
\begin{equation}
\begin{aligned}
A_{1,\xi } = \frac{d-2}{d-1}B \sum_\chi \frac{D^{(c)}_{\xi,\chi} }{F_\chi^2 M_{6^*}} \frac{2Q_\xi}{\omega}\mathcal{G}_2(\omega,\delta,M_\chi^2),
\end{aligned}
\end{equation}
\begin{equation}
\begin{aligned}
A_{3,\xi } =& -\frac{1}{d-1}B \sum_\chi \frac{D^{(c)}_{\xi,\chi} }{F_\chi^2 M_{6^*}}\frac{2Q_\xi}{\omega} \\
&\left[\mathcal{J}_2(\omega,\delta,M_\chi^2) - \mathcal{J}_2(0,\delta,M_\chi^2)\right].
\end{aligned}
\end{equation}

For diagram $(d'_2+e'_2)$:
\begin{equation}
\begin{aligned}
A_{1,\xi} =& \frac{d-2}{d-1}B \sum_\chi \frac{D^{(d+e)}_{\xi,\chi}}{2M_{6^*} F^2_\chi}\left\{2I_2(M_\chi^2) + \int_0^1 dx \frac{1}{1-x} \right. \\
& \left\{\frac{\partial}{\partial \omega}\mathcal{G}_7^\prime((1-x)\omega,\delta,M_\chi^2)\right. \\
&+(d+2)\frac{\partial}{\partial \omega}\mathcal{G}_6^\prime((1-x)\omega,\delta,M_\chi^2) \\
& +2\omega(1-x)\frac{\partial}{\partial \omega}\mathcal{J}_4^\prime((1-x)\omega,\delta,M_\chi^2) \\
& -2\delta \frac{\partial}{\partial \omega}\mathcal{G}_4^\prime((1-x)\omega,\delta,M_\chi^2) \\
& +[\delta^2-\omega^2(1-\cos\theta)x]\frac{\partial}{\partial \omega}\mathcal{G}_2^\prime((1-x)\omega,\delta,M_\chi^2) \\
&\left.\left. +2\omega \delta(x-1)\frac{\partial}{\partial \omega}\mathcal{J}_2^\prime((1-x)\omega,\delta,M_\chi^2)\right\}\right\},
\end{aligned}
\label{eqc.70}
\end{equation}
\begin{equation}
\begin{aligned}
A_{2,\xi} =& -\frac{d-2}{d-1}B \sum_\chi \frac{D^{(d+e)}_{\xi,\chi}}{2M_{6^*} F^2_\chi}\int_0^1 dx \omega^2 \\
&\frac{\partial}{\partial \omega}\mathcal{G}_2^\prime((1-x)\omega,\delta,M_\chi^2),
\end{aligned}
\label{eqc.71}
\end{equation}
\begin{equation}
\begin{aligned}
A_{3,\xi} =& -\frac{1}{d-1}B \sum_\chi \frac{D^{(d+e)}_{\xi,\chi}}{2M_{6^*} F^2_\chi}\int_0^1 dx \frac{1}{1-x} \\
& \left\{\frac{\partial}{\partial \omega}\mathcal{J}_7^\prime((1-x)\omega,\delta,M_\chi^2)\right. \\
& +(d+2)\frac{\partial}{\partial \omega}\mathcal{J}_6^\prime((1-x)\omega,\delta,M_\chi^2) \\
& +2\omega(1-x)\frac{\partial}{\partial \omega}\mathcal{G}_4^\prime((1-x)\omega,\delta,M_\chi^2) \\
& -2\delta \frac{\partial}{\partial \omega}\mathcal{J}_4^\prime((1-x)\omega,\delta,M_\chi^2) \\
& +[\delta^2-\omega^2(1-\cos\theta)x]\frac{\partial}{\partial \omega}\mathcal{J}_2^\prime((1-x)\omega,\delta,M_\chi^2) \\
&\left. +2\omega \delta (x-1)\frac{\partial}{\partial \omega}\mathcal{G}_2^\prime((1-x)\omega,\delta,M_\chi^2)\right\},
\end{aligned}
\label{eqc.72}
\end{equation}
\begin{equation}
\begin{aligned}
A_{5,\xi} =& -\frac{1}{d-1}B \sum_\chi \frac{D^{(d+e)}_{\xi,\chi}}{2M_{6^*} F^2_\chi} \omega^2 \int_0^1 dx (-1) \\
& \frac{\partial}{\partial \omega}J_2^\prime(-(1-x)\omega - \delta,M_\chi^2),
\end{aligned}
\label{eqc.73}
\end{equation}

For diagram $(d'_3+e'_3)$:
\begin{equation}
\begin{aligned}
A_{1,\xi} =& \frac{d-2}{d-1}B \sum_\chi \frac{D^{(d_3+e_3)}_{\xi,\chi} }{F_\chi^2 M_6} \left\{I_2(M_\chi^2) + \int_0^1 dx \right. \\
&\left. \left[-\frac{\omega}{2} \mathcal{G}^\prime_2(x\omega,\delta,M^2_\chi) + \frac{\delta}{2} \mathcal{J}^\prime_2(x\omega,\delta,M^2_\chi)\right]\right\},
\end{aligned}
\label{eqc.75}
\end{equation}

\begin{equation}
\begin{aligned}
A_{3,\xi} =& -\frac{1}{d-1}B \sum_\chi \frac{D^{(d_3+e_3)}_{\xi,\chi} }{F_\chi^2 M_6} \int_0^1 dx \\
&\left[-\frac{\omega}{2} \mathcal{J}^\prime_2(x\omega,\delta,M^2_\chi) + \frac{\delta}{2} \mathcal{G}^\prime_2(x\omega,\delta,M^2_\chi)\right],
\end{aligned}
\label{eqc.77}
\end{equation}

For diagram $(d'_4+e'_4)$:
\begin{equation}
\begin{aligned}
A_{1,\xi} =& \frac{d-2}{d-1}B \sum_\chi \frac{D^{(d+e)}_{\xi,\chi}}{2F_\chi^2 M_6}\left[-2 I_2(M^2_\chi) - \int_0^1 dx \right. \\
&\left.\delta \mathcal{J}^\prime_2(x\omega , \delta,M^2_\chi)\right],
\end{aligned}
\label{eqc.79}
\end{equation}

\begin{equation}
\begin{aligned}
A_{3,\xi} =& -\frac{1}{d-1}B \sum_\chi \frac{D^{(d+e)}_{\xi,\chi}}{2F_\chi^2 M_6}\int_0^1 dx (-\delta) \mathcal{G}^\prime_2(x\omega , \delta,M^2_\chi),
\end{aligned}
\label{eqc.81}
\end{equation}

For diagram $(d'_5+e'_5)$:
\begin{equation}
\begin{aligned}
A_{1,\xi} =& -\frac{1}{d-1}B \sum_\chi \frac{D^{(d+e)}_{\xi,\chi} C_{2,\xi}}{2M_N F^2_\chi}\int_0^1 dx \left(-\omega \cos\theta \right)\\
& \mathcal{G}^\prime_2\left((1-x)\omega,\delta,M^2_\chi\right),
\end{aligned}
\label{eqc.84}
\end{equation}
\begin{equation}
\begin{aligned}
A_{2,\xi} =& -\frac{1}{d-1}B \sum_\chi \frac{D^{(d+e)}_{\xi,\chi} C_{2,\xi}}{2M_N F^2_\chi}\int_0^1 dx \omega \\
&\mathcal{G}^\prime_2\left((1-x)\omega,\delta,M^2_\chi\right),
\end{aligned}
\label{eqc.85}
\end{equation}
\begin{equation}
\begin{aligned}
A_{3,\xi} =& -\frac{1}{d-1}B \sum_\chi \frac{D^{(d+e)}_{\xi,\chi} C_{2,\xi}}{2M_N F^2_\chi}\int_0^1 dx \left(-\omega \cos\theta \right)\\
&\mathcal{J}^\prime_2\left((1-x)\omega,\delta,M^2_\chi\right),
\end{aligned}
\label{eqc.86}
\end{equation}
\begin{equation}
\begin{aligned}
A_{4,\xi} =& -\frac{1}{d-1}B \sum_\chi \frac{D^{(d+e)}_{\xi,\chi} C_{2,\xi}}{2M_N F^2_\chi}\int_0^1 dx \left(-\omega \right)\\
&\mathcal{J}^\prime_2\left((1-x)\omega,\delta,M^2_\chi\right),
\end{aligned}
\label{eqc.87}
\end{equation}
\begin{equation}
\begin{aligned}
A_{5,\xi} =& -\frac{1}{d-1}B \sum_\chi \frac{D^{(d+e)}_{\xi,\chi} C_{2,\xi}}{2M_N F^2_\chi}\int_0^1 dx \omega \\
& \mathcal{J}^\prime_2\left((1-x)\omega,\delta,M^2_\chi\right).
\end{aligned}
\label{eqc.88}
\end{equation}

\begin{equation}
\begin{aligned}
A_{6,\xi} =& \frac{1}{d-1}B \sum_\chi \frac{D^{(d+e)}_{\xi,\chi} Q_\xi}{2M_6 F^2_\chi}\int_0^1 dx \omega \\
& J^\prime_2\left(-(1-x)\omega - \delta,M^2_\chi\right).
\end{aligned}
\end{equation}

For diagram $(d''_5+e''_5)$:
\begin{equation}
\begin{aligned}
A_{1,\xi} =& -\frac{1}{d-1}\frac{e^2 g_1 g_3}{2} \sum_\chi \frac{D^{(d+e)}_{\xi,\chi} C_{1,\xi}}{4M_N F^2_\chi}\int_0^1 dx \omega^2 \cos\theta\\
& \left[\frac{1}{\omega - \delta} J^\prime_2\left((1-x)\omega , M^2_\chi\right)\right. \\
&\left. - \frac{1}{\omega + \delta} J^\prime_2\left(-(1-x)\omega , M^2_\chi\right)\right],
\end{aligned}
\label{eqc.89}
\end{equation}
\begin{equation}
\begin{aligned}
A_{2,\xi} =& -\frac{1}{d-1}\frac{e^2 g_1 g_3}{2} \sum_\chi \frac{D^{(d+e)}_{\xi,\chi} C_{1,\xi}}{4M_N F^2_\chi}\int_0^1 dx \omega^2 \\
&\left[-\frac{1}{\omega - \delta}J^\prime_2\left((1-x)\omega ,M^2_\chi\right)\right. \\
&\left. + \frac{1}{\omega + \delta}J^\prime_2\left(-(1-x)\omega ,M^2_\chi\right)\right],
\end{aligned}
\label{eqc.90}
\end{equation}
\begin{equation}
\begin{aligned}
A_{3,\xi} =& -\frac{1}{d-1}\frac{e^2 g_1 g_3}{2} \sum_\chi \frac{D^{(d+e)}_{\xi,\chi} C_{1,\xi}}{4M_N F^2_\chi}\int_0^1 dx \omega^2 \cos\theta\\
&\left[\frac{1}{\omega - \delta} (3x-2)J^\prime_2\left((1-x)\omega ,M^2_\chi\right)\right. \\
&\left. + \frac{1}{\omega + \delta} (3x-2)J^\prime_2\left(-(1-x)\omega ,M^2_\chi\right)\right],
\end{aligned}
\label{eqc.91}
\end{equation}
\begin{equation}
\begin{aligned}
A_{4,\xi} =& -\frac{1}{d-1}\frac{e^2 g_1 g_3}{2} \sum_\chi \frac{D^{(d+e)}_{\xi,\chi} C_{1,\xi}}{4M_N F^2_\chi}\int_0^1 dx \omega^2 \\
&\left[\frac{1}{\omega - \delta}(1-3x)J^\prime_2\left((1-x)\omega ,M^2_\chi\right)\right. \\
&\left. + \frac{1}{\omega + \delta}(1-3x)J^\prime_2\left(-(1-x)\omega ,M^2_\chi\right)\right],
\end{aligned}
\label{eqc.92}
\end{equation}
\begin{equation}
\begin{aligned}
A_{5,\xi} =& -\frac{1}{d-1}\frac{e^2 g_1 g_3}{2} \sum_\chi \frac{D^{(d+e)}_{\xi,\chi} C_{1,\xi}}{4M_N F^2_\chi}\int_0^1 dx \omega^2 \\
&\left[\frac{1}{2(\omega - \delta)} J^\prime_2\left((1-x)\omega ,M^2_\chi\right)\right. \\
&\left. + \frac{1}{2(\omega + \delta)} J^\prime_2\left(-(1-x)\omega ,M^2_\chi\right)\right].
\end{aligned}
\label{eqc.93}
\end{equation}

For diagram $(d'''_5+e'''_5)$:
\begin{equation}
\begin{aligned}
A_{1,\xi} =& \frac{1}{(d-1)^2}\frac{e^2 g_3 g_5}{2} \sum_\chi \frac{D^{(d+e)}_{\xi,\chi} C_{1,\xi}}{4M_N F^2_\chi}\int_0^1 dx \omega^2 \cos\theta\\
&\left[-\frac{5}{\omega - \delta}J^\prime_2\left((1-x)\omega -\delta,M^2_\chi\right)\right. \\
&\left. + \frac{5}{\omega + \delta}J^\prime_2\left(-(1-x)\omega -\delta,M^2_\chi\right)\right],
\end{aligned}
\label{eqc.94}
\end{equation}
\begin{equation}
\begin{aligned}
A_{2,\xi} =& \frac{1}{(d-1)^2}\frac{e^2 g_3 g_5}{2} \sum_\chi \frac{D^{(d+e)}_{\xi,\chi} C_{1,\xi}}{4M_N F^2_\chi}\int_0^1 dx \omega^2 \\
&\left[\frac{5}{\omega - \delta}J^\prime_2\left((1-x)\omega -\delta,M^2_\chi\right)\right. \\
&\left. - \frac{5}{\omega + \delta}J^\prime_2\left(-(1-x)\omega -\delta,M^2_\chi\right)\right],
\end{aligned}
\label{eqc.95}
\end{equation}
\begin{equation}
\begin{aligned}
A_{3,\xi} =& \frac{1}{(d-1)^2}\frac{e^2 g_3 g_5}{2} \sum_\chi \frac{D^{(d+e)}_{\xi,\chi} C_{1,\xi}}{4M_N F^2_\chi}\int_0^1 dx \omega^2 \cos\theta\\
&\left[\frac{1}{\omega - \delta}(3x+1)J^\prime_2\left((1-x)\omega -\delta,M^2_\chi\right)\right. \\
&\left. + \frac{1}{\omega + \delta}(3x+1)J^\prime_2\left(-(1-x)\omega -\delta,M^2_\chi\right)\right],
\end{aligned}
\label{eqc.96}
\end{equation}
\begin{equation}
\begin{aligned}
A_{4,\xi} =& \frac{1}{(d-1)^2}\frac{e^2 g_3 g_5}{2} \sum_\chi \frac{D^{(d+e)}_{\xi,\chi} C_{1,\xi}}{4M_N F^2_\chi}\int_0^1 dx \omega^2\\
&\left[\frac{1}{\omega - \delta}(-3x+4)J^\prime_2\left((1-x)\omega -\delta,M^2_\chi\right)\right. \\
&\left. + \frac{1}{\omega + \delta}(-3x+4)J^\prime_2\left(-(1-x)\omega -\delta,M^2_\chi\right)\right],
\end{aligned}
\label{eqc.97}
\end{equation}
\begin{equation}
\begin{aligned}
A_{5,\xi} =& \frac{1}{(d-1)^2}\frac{e^2 g_3 g_5}{2} \sum_\chi \frac{D^{(d+e)}_{\xi,\chi} C_{1,\xi}}{4M_N F^2_\chi}\int_0^1 dx \omega^2 \\
&\left[-\frac{5}{2(\omega - \delta)}J^\prime_2\left((1-x)\omega -\delta,M^2_\chi\right)\right. \\
&\left. - \frac{5}{2(\omega + \delta)} J^\prime_2\left(-(1-x)\omega -\delta,M^2_\chi\right)\right].
\end{aligned}
\label{eqc.98}
\end{equation}

For diagram $f'_2$:
\begin{equation}
\begin{aligned}
A_{1,\xi} =& \frac{d-2}{d-1}B \sum_\chi \frac{D^{(f)}_{\xi ,\chi}}{2F_\chi^2 M_{6^*}}\int_0^1 dx \left\{\int_0^{1-x} dy \right.  \\
& \left\{(d+1) \left[2I_6^\prime(\mathcal{M}_\chi) - \left(M^2_\chi - \delta^2\right) \right.\right. \\
&\frac{\partial}{\partial a}\mathcal{G}_6^{\prime \prime}((1-x-y)\omega - a,\delta,\mathcal{M}_\chi) \\
&\left. + 2\delta \mathcal{J}_6^{\prime \prime}((1-x-y)\omega ,\delta,\mathcal{M}_\chi)\right] - \omega^2 \\
& \left[x^2+y^2+2xy\cos\theta - (x+y)(1+\cos\theta)+\cos\theta\right] \\
& \left[2I_2^\prime(\mathcal{M}_\chi) -\left(M^2_\chi - \delta^2\right) \right. \\
&\frac{\partial}{\partial a}\mathcal{G}_2^{\prime \prime}((1-x-y)\omega - a,\delta,\mathcal{M}_\chi) \\
&\left.\left. + 2\delta \mathcal{J}_2^{\prime \prime}((1-x-y)\omega ,\delta,\mathcal{M}_\chi)\right]\right\} \\
& + \left\{\frac{d+1}{1-x}\frac{\partial}{\partial \omega}\mathcal{G}_6^\prime((1-x)\omega,\delta,M^2_\chi)\right. \\
& - \frac{1}{1-x}\omega^2\left[x^2 - x(1+\cos\theta)+\cos\theta\right] \\
&\left.\left. \frac{\partial}{\partial \omega}\mathcal{G}_2^\prime((1-x)\omega,\delta,M^2_\chi)\right\}\right\},
\end{aligned}
\label{eqc.99}
\end{equation}
\begin{equation}
\begin{aligned}
A_{2,\xi} =& \frac{d-2}{d-1}B \sum_\chi \frac{D^{(f)}_{\xi ,\chi}}{2F_\chi^2 M_{6^*}} \int_0^1 dx \omega^2\left\{ \int_0^{1-x} dy \right. \\
& \left[-(3+d)xy + x + y\right] \\
& \left[2I_2^\prime(\mathcal{M}_\chi) - \left(M^2_\chi - \delta^2\right)\right. \\
& \frac{\partial}{\partial a}\mathcal{G}_2^{\prime \prime}((1-x-y)\omega - a,\delta,\mathcal{M}_\chi) \\
&\left. + 2\delta \mathcal{J}_2^{\prime \prime}((1-x-y)\omega,\delta,\mathcal{M}_\chi)\right] \\
& + \frac{x}{1-x} \left.\frac{\partial}{\partial \omega}\mathcal{G}^\prime_2((1-x)\omega,\delta,M^2_\chi) \right\},
\end{aligned}
\label{eqc.100}
\end{equation}
\begin{equation}
\begin{aligned}
A_{4,\xi} =& -\frac{1}{d-1}B \sum_\chi \frac{D^{(f)}_{\xi ,\chi}}{2F_\chi^2 M_{6^*}}\int_0^1 dx \omega^2\left[ \int_0^{1-x} dy \right. \\
& (-x-y+1)\left[\left(M^2_\chi - \delta^2\right) \frac{\partial}{\partial a}\right. \\
& \mathcal{J}_2^{\prime \prime}((1-x-y)\omega - a,\delta,\mathcal{M}_\chi) \\
& \left. - 2\delta \mathcal{G}_2^{\prime \prime}((1-x-y)\omega ,\delta,\mathcal{M}_\chi)\right] \\
&\left. - \frac{\partial}{\partial \omega}\mathcal{J}^\prime_2((1-x)\omega,\delta,M^2_\chi)\right],
\end{aligned}
\label{eqc.101}
\end{equation}
\begin{equation}
\begin{aligned}
A_{5,\xi} =& -\frac{1}{d-1}B \sum_\chi \frac{D^{(f)}_{\xi ,\chi}}{2F_\chi^2 M_{6^*}} \int_0^1 dx \omega^2\left[\int_0^{1-x} dy \right. \\
& (-x)\left[\left(M^2_\chi - \delta^2\right) \frac{\partial}{\partial a}\right. \\
& \mathcal{J}_2^{\prime \prime}((1-x-y)\omega - a,\delta,\mathcal{M}_\chi) \\
&\left. - 2\delta \mathcal{G}_2^{\prime \prime}((1-x-y)\omega,\delta,\mathcal{M}_\chi)\right] \\
& + \frac{x}{1-x} \frac{\partial}{\partial \omega}J^\prime_2((1-x)\omega-\delta,M^2_\chi)\left.\right],
\end{aligned}
\label{eqc.102}
\end{equation}
\begin{equation}
\begin{aligned}
A_{6,\xi} =& -\frac{1}{d-1}B \sum_\chi \frac{D^{(f)}_{\xi ,\chi}}{2F_\chi^2 M_{6^*}} \int_0^1 dx \omega^2\left[\int_0^{1-x} dy \right. \\
& x\left[ \left(M^2_\chi - \delta^2\right) \frac{\partial}{\partial a}\right. \\
& \mathcal{J}_2^{\prime \prime}((1-x-y)\omega - a,\delta,\mathcal{M}_\chi)\\
&\left. - 2\delta \mathcal{G}_2^{\prime \prime}((1-x-y)\omega ,\delta,\mathcal{M}_\chi)\right] \\
& - \frac{x}{1-x} \frac{\partial}{\partial \omega}J^\prime_2((1-x)\omega-\delta,M^2_\chi)\left. \right].
\end{aligned}
\label{eqc.103}
\end{equation}

For diagram $f'_3$:
\begin{equation}
\begin{aligned}
A_{1,\xi} =& \frac{d-2}{d-1}B \sum_\chi \frac{D^{(f)}_{\xi ,\chi}}{2F_\chi^2 M_6}\int_0^1 dx \int_0^{1-x} dy \left\{-2(d+1)\right. \\
& \left[2I_6^\prime(\mathcal{M}_\chi) + \delta \mathcal{J}^{\prime \prime}_6((1-x-y)\omega,\delta,\mathcal{M}_\chi)\right] + 2\omega^2 \\
& \left[(x^2+y^2+2xy\cos\theta)-(x + y)(1 + \cos\theta)+1\right] \\
& \left. \left[2I_2^\prime(\mathcal{M}_\chi) + \delta \mathcal{J}^{\prime \prime}_2((1-x-y)\omega,\delta,\mathcal{M}_\chi)\right]\right\},
\end{aligned}
\label{eqc.104}
\end{equation}
\begin{equation}
\begin{aligned}
A_{2,\xi} =& \frac{d-2}{d-1}B \sum_\chi \frac{D^{(f)}_{\xi ,\chi}}{2F_\chi^2 M_6} \int_0^1 dx \int_0^{1-x} dy \\
& 2\omega^2 \left[x+y-(d+3)xy\right] \\
& \left[2I_2^\prime(\mathcal{M}_\chi) + \delta \mathcal{J}^{\prime \prime}_2((1-x-y)\omega,\delta,\mathcal{M}_\chi)\right].
\end{aligned}
\label{eqc.105}
\end{equation}

For diagram $f'_4$:
\begin{equation}
\begin{aligned}
A_{1,\xi} =& -\frac{1}{d-1}\frac{e^2 g_1 g_3}{2} \sum_\chi \frac{D^{(f)}_{\xi,\chi} C_{1,\xi}}{4M_N F^2_\chi}\int_0^1 dx \\
&\left\{-\frac{1}{2(\omega - \delta)}\omega^2 \cos\theta \left[J^\prime_2\left((1-x)\omega-\delta,M^2_\chi\right) \right.\right.\\
&\left. -J^\prime_2\left((-x)\omega,M^2_\chi\right)\right]+\frac{1}{2(\omega + \delta)}\omega^2 \cos\theta \\
&\left. \left[J^\prime_2\left(-(1-x)\omega-\delta,M^2_\chi\right) -J^\prime_2\left(-(-x)\omega,M^2_\chi\right)\right]\right\},
\end{aligned}
\label{eqc.106}
\end{equation}
\begin{equation}
\begin{aligned}
A_{2,\xi} =& -\frac{1}{d-1}\frac{e^2 g_1 g_3}{2} \sum_\chi \frac{D^{(f)}_{\xi,\chi} C_{1,\xi}}{4M_N F^2_\chi}\int_0^1 dx \\
&\left\{\frac{1}{2(\omega - \delta)}\omega^2 \left[J^\prime_2\left((1-x)\omega-\delta,M^2_\chi\right)\right.\right.\\
&\left.-J^\prime_2\left((-x)\omega,M^2_\chi\right)\right]-\frac{1}{2(\omega + \delta)}\omega^2 \\
&\left. \left[J^\prime_2\left(-(1-x)\omega-\delta,M^2_\chi\right)-J^\prime_2\left(-(-x)\omega,M^2_\chi\right)\right]\right\},
\end{aligned}
\label{eqc.107}
\end{equation}
\begin{equation}
\begin{aligned}
A_{3,\xi} =& -\frac{1}{d-1}\frac{e^2 g_1 g_3}{2} \sum_\chi \frac{D^{(f)}_{\xi,\chi} C_{1,\xi}}{4M_N F^2_\chi}\int_0^1 dx \\
&\left\{\frac{x+1}{2(\omega - \delta)} \omega^2 \cos\theta \left[J^\prime_2\left((1-x)\omega-\delta,M^2_\chi\right)\right.\right.\\
&\left.-J^\prime_2\left((-x)\omega,M^2_\chi\right)\right]+\frac{x+1}{2(\omega + \delta)} \omega^2 \cos\theta \\
&\left. \left[J^\prime_2\left(-(1-x)\omega-\delta,M^2_\chi\right)-J^\prime_2\left(-(-x)\omega,M^2_\chi\right)\right]\right\},
&\end{aligned}
\label{eqc.108}
\end{equation}
\begin{equation}
\begin{aligned}
A_{4,\xi} =& -\frac{1}{d-1}\frac{e^2 g_1 g_3}{2} \sum_\chi \frac{D^{(f)}_{\xi,\chi} C_{1,\xi}}{4M_N F^2_\chi}\int_0^1 dx \\
&\left\{-\frac{x-2}{2(\omega - \delta)} \omega^2 \left[J^\prime_2\left((1-x)\omega-\delta,M^2_\chi\right) \right.\right.\\
&\left.-J^\prime_2\left((-x)\omega,M^2_\chi\right)\right]-\frac{x-2}{2(\omega + \delta)} \omega^2 \\
&\left. \left[J^\prime_2\left(-(1-x)\omega-\delta,M^2_\chi\right)-J^\prime_2\left(-(-x)\omega,M^2_\chi\right)\right]\right\},
\end{aligned}
\label{eqc.109}
\end{equation}
\begin{equation}
\begin{aligned}
A_{5,\xi} =& -\frac{1}{d-1}\frac{e^2 g_1 g_3}{2} \sum_\chi \frac{D^{(f)}_{\xi,\chi} C_{1,\xi}}{4M_N F^2_\chi}\int_0^1 dx \\
&\left\{\frac{x-2}{2(\omega - \delta)}\omega^2 \left[J^\prime_2\left((1-x)\omega-\delta,M^2_\chi\right)\right.\right.\\
&\left.-J^\prime_2\left((-x)\omega,M^2_\chi\right)\right]-\frac{x+1}{2(\omega + \delta)}\omega^2 \\
&\left. \left[J^\prime_2\left(-(1-x)\omega-\delta,M^2_\chi\right)-J^\prime_2\left(-(-x)\omega,M^2_\chi\right)\right]\right\}.
\end{aligned}
\label{eqc.110}
\end{equation}

For diagram $f''_4$:
\begin{equation}
\begin{aligned}
A_{1,\xi} =& -\frac{1}{d-1}\frac{e^2 g_1 g_3}{2} \sum_\chi \frac{D^{(f)}_{\xi,\chi} C_{1,\xi}}{4M_N F^2_\chi}\int_0^1 dx \\
&\left\{-\frac{1}{2(\omega + \delta)}\omega^2 \cos\theta \left[J^\prime_2\left((1-x)\omega,M^2_\chi\right) \right.\right.\\
&\left. -J^\prime_2\left((-x)\omega-\delta,M^2_\chi\right)\right]+\frac{1}{2(\omega - \delta)}\omega^2 \cos\theta \\
&\left. \left[J^\prime_2\left(-(1-x)\omega,M^2_\chi\right) -J^\prime_2\left(-(-x)\omega-\delta,M^2_\chi\right)\right]\right\},
\end{aligned}
\label{eqc.111}
\end{equation}
\begin{equation}
\begin{aligned}
A_{2,\xi} =& -\frac{1}{d-1}\frac{e^2 g_1 g_3}{2} \sum_\chi \frac{D^{(f)}_{\xi,\chi} C_{1,\xi}}{4M_N F^2_\chi}\int_0^1 dx \\
&\left\{\frac{1}{2(\omega + \delta)}\omega^2 \left[J^\prime_2\left((1-x)\omega,M^2_\chi\right)\right.\right.\\
&\left.-J^\prime_2\left((-x)\omega-\delta,M^2_\chi\right)\right]-\frac{1}{2(\omega - \delta)}\omega^2 \\
&\left. \left[J^\prime_2\left(-(1-x)\omega,M^2_\chi\right)-J^\prime_2\left(-(-x)\omega-\delta,M^2_\chi\right)\right]\right\},
\end{aligned}
\label{eqc.112}
\end{equation}
\begin{equation}
\begin{aligned}
A_{3,\xi} =& -\frac{1}{d-1}\frac{e^2 g_1 g_3}{2} \sum_\chi \frac{D^{(f)}_{\xi,\chi} C_{1,\xi}}{4M_N F^2_\chi}\int_0^1 dx \\
&\left\{\frac{x-2}{2(\omega + \delta)} \omega^2 \cos\theta \left[J^\prime_2\left((1-x)\omega,M^2_\chi\right)\right.\right.\\
&\left.-J^\prime_2\left((-x)\omega-\delta,M^2_\chi\right)\right]+\frac{x-2}{2(\omega - \delta)} \omega^2 \cos\theta \\
&\left. \left[J^\prime_2\left(-(1-x)\omega,M^2_\chi\right)-J^\prime_2\left(-(-x)\omega-\delta,M^2_\chi\right)\right]\right\},
\end{aligned}
\label{eqc.113}
\end{equation}
\begin{equation}
\begin{aligned}
A_{4,\xi} =& -\frac{1}{d-1}\frac{e^2 g_1 g_3}{2} \sum_\chi \frac{D^{(f)}_{\xi,\chi} C_{1,\xi}}{4M_N F^2_\chi}\int_0^1 dx \\
&\left\{-\frac{x+1}{2(\omega + \delta)} \omega^2 \left[J^\prime_2\left((1-x)\omega,M^2_\chi\right) \right.\right.\\
&\left.-J^\prime_2\left((-x)\omega-\delta,M^2_\chi\right)\right]-\frac{x+1}{2(\omega - \delta)} \omega^2 \\
&\left. \left[J^\prime_2\left(-(1-x)\omega,M^2_\chi\right)-J^\prime_2\left(-(-x)\omega-\delta,M^2_\chi\right)\right]\right\},
\end{aligned}
\label{eqc.114}
\end{equation}
\begin{equation}
\begin{aligned}
A_{5,\xi} =& -\frac{1}{d-1}\frac{e^2 g_1 g_3}{2} \sum_\chi \frac{D^{(f)}_{\xi,\chi} C_{1,\xi}}{4M_N F^2_\chi}\int_0^1 dx \\
&\left\{\frac{x+1}{2(\omega + \delta)}\omega^2 \left[J^\prime_2\left((1-x)\omega,M^2_\chi\right)\right.\right.\\
&\left.-J^\prime_2\left((-x)\omega-\delta,M^2_\chi\right)\right]-\frac{x-2}{2(\omega - \delta)}\omega^2 \\
&\left. \left[J^\prime_2\left(-(1-x)\omega,M^2_\chi\right)-J^\prime_2\left(-(-x)\omega-\delta,M^2_\chi\right)\right]\right\}.
\end{aligned}
\label{eqc.115}
\end{equation}

For diagram $f'''_4$:
\begin{equation}
\begin{aligned}
A_{1,\xi} =& \frac{1}{(d-1)^2}B \sum_\chi \frac{D^{(f)}_{\xi,\chi} \tilde{C}_{3,\xi,\chi}}{4M_N F^2_\chi}\int_0^1 dx 5\omega \cos\theta \\
& \left[\mathcal{G}^\prime_2\left((1-x)\omega,\delta,M^2_\chi\right)-\mathcal{G}^\prime_2\left((-x)\omega ,\delta,M^2_\chi\right)\right] \\
& + \frac{d-2}{d-1}B \sum_\chi \frac{D^{(f)}_{\xi,\chi} Q_\xi}{2M^*_6 F^2_\chi}\int_0^1 dx \left\{\left(-\frac{d+1}{\omega}\right)\right. \\
& \left[\mathcal{G}^\prime_6\left((1-x)\omega,\delta,M^2_\chi\right)-\mathcal{G}^\prime_6\left((-x)\omega ,\delta,M^2_\chi\right)\right] \\
&+ \omega x(x-1) \left[\mathcal{G}^\prime_2\left((1-x)\omega,\delta,M^2_\chi\right) \right. \\
&\left. \left.-\mathcal{G}^\prime_2\left((-x)\omega ,\delta,M^2_\chi\right)\right]\right\},
\end{aligned}
\label{eqc.116}
\end{equation}
\begin{equation}
\begin{aligned}
A_{2,\xi} =& \frac{1}{(d-1)^2}B \sum_\chi \frac{D^{(f)}_{\xi,\chi} \tilde{C}_{3,\xi,\chi}}{4M_N F^2_\chi}\int_0^1 dx \left(-5\omega \right) \\
& \left[\mathcal{G}^\prime_2\left((1-x)\omega,\delta,M^2_\chi\right)-\mathcal{G}^\prime_2\left((-x)\omega ,\delta,M^2_\chi\right)\right],
\end{aligned}
\label{eqc.117}
\end{equation}
\begin{equation}
\begin{aligned}
A_{3,\xi} =& \frac{1}{(d-1)^2}B \sum_\chi \frac{D^{(f)}_{\xi,\chi} \tilde{C}_{3,\xi,\chi}}{4M_N F^2_\chi}\int_0^1 dx (1-2x) \omega \cos\theta  \\
&\left[\mathcal{J}^\prime_2\left((1-x)\omega ,\delta,M^2_\chi\right)-\mathcal{J}^\prime_2\left((-x)\omega ,\delta,M^2_\chi\right)\right],
\end{aligned}
\label{eqc.118}
\end{equation}
\begin{equation}
\begin{aligned}
A_{4,\xi} =& \frac{1}{(d-1)^2}B \sum_\chi \frac{D^{(f)}_{\xi,\chi} \tilde{C}_{3,\xi,\chi}}{4M_N F^2_\chi}\int_0^1 dx (2x-1)\omega \\
&\left[\mathcal{J}^\prime_2\left((1-x)\omega ,\delta,M^2_\chi\right)-\mathcal{J}^\prime_2\left((-x)\omega ,\delta,M^2_\chi\right)\right],
\end{aligned}
\label{eqc.119}
\end{equation}
\begin{equation}
\begin{aligned}
A_{5,\xi} =& \frac{1}{(d-1)^2}B \sum_\chi \frac{D^{(f)}_{\xi,\chi} \tilde{C}_{3,\xi,\chi}}{4M_N F^2_\chi}\int_0^1 dx (1-2x)\omega  \\
&\left[\mathcal{G}^\prime_2\left((1-x)\omega ,\delta,M^2_\chi\right)-\mathcal{G}^\prime_2\left((-x)\omega ,\delta,M^2_\chi\right)\right].
\end{aligned}
\label{eqc.120}
\end{equation}

\begin{equation}
\begin{aligned}
A_{6,\xi} =& -\frac{1}{d-1}B \sum_\chi \frac{D^{(f)}_{\xi,\chi} Q_\xi}{2M^*_6 F^2_\chi}\int_0^1 dx \omega(x-1)  \\
&\left[J^\prime_2\left(-(1-x)\omega -\delta,M^2_\chi\right) \right. \\
&\left.-J^\prime_2\left(-(-x)\omega -\delta,M^2_\chi\right)\right].
\end{aligned}
\end{equation}

For diagram $g'_2$:
\begin{equation}
\begin{aligned}
A_{1,\xi} =& \frac{d-2}{d-1}B \sum_\chi \frac{D^{(g)}_{\xi,\chi}}{2M_6 F^2_\chi} \int_0^1 dx \left[2(d-1)I_2(\mathcal{M}^\prime_\chi)\right. \\
& + 4x(1-x)\omega^2 (1-\cos\theta)I_0(\mathcal{M}^\prime_\chi) \\
& - 2(d-1)\frac{\partial}{\partial a}J^\prime_7(-a-\delta,\mathcal{M}^\prime_\chi) \\
& - 4(d+5) \frac{\partial}{\partial a}J^\prime_6(-a-\delta,\mathcal{M}^\prime_\chi) \\
& + 4x(x-1)\omega^2(1-\cos\theta)\frac{\partial}{\partial a}J^\prime_3(-a-\delta,\mathcal{M}^\prime_\chi) \\
&\left. + x\left(44x-24\right)\omega^2(1-\cos\theta)\frac{\partial}{\partial a}J^\prime_2(-a-\delta,\mathcal{M}^\prime_\chi)\right].
\end{aligned}
\label{eqc.121}
\end{equation}

For diagram $g'_3$:
\begin{equation}
\begin{aligned}
A_{1,\xi} =& \frac{d-2}{d-1}B \sum_\chi \frac{D^{(g)}_{\xi,\chi}}{M_6 F^2_\chi}\int_0^1 dx \left[(d-1)\left(2I_2(\mathcal{M}^\prime_\chi) \right.\right. \\
&\left. + \delta \mathcal{J}'_2(-a,\delta,\mathcal{M}^\prime_\chi)\right) + 2 \omega^2 x^2 \\
& \left.(\cos\theta - 1)\left(2I_0(\mathcal{M}^\prime_\chi) + \delta \mathcal{J}'_0(-a,\delta,\mathcal{M}^\prime_\chi)\right)\right].
\end{aligned}
\label{eqc.122}
\end{equation}

\nocite{*}

\bibliography{polarizabilities}

\end{document}